\documentclass[apj]{emulateapj}
\usepackage{amsmath} 
\usepackage{natbib}
\usepackage{hyperref}
\usepackage{amssymb}
\usepackage{epsfig}
\usepackage{apjfonts}
\usepackage{graphicx}
\usepackage{subfigure}
\usepackage{afterpage}
\newcommand{\cms}{\,{\rm cm$^{-2}$}\,}

\newcommand{\kms}{\,{\rm km\,s$^{-1}$}\,} 
\newcommand{\kmsmpc}{\,{\rm km\,s$^{-1}$\,Mpc$^{-1}$}\,}

\newcommand{\photflux}{\,{\rm photons\,s$^{-1}$\,cm$^{-2}$}\,}

\newcommand{\ergs}{\,{\rm erg\,s$^{-1}$}\,}
\newcommand{\ergscm}{\,{\rm erg\,s$^{-1}$\,cm$^{-2}$}\,}
\newcommand{\Ms}{M_\odot}
\newcommand{\Zs}{Z_\odot}

\shorttitle{Kreisch et al: RXJ1347.5-1145 Hydrodynamics}

\begin{document}


\title{Merger Hydrodynamics of the Luminous Cluster RXJ1347.5-1145}

\author{C.D. Kreisch$^{1,2}$, M.E. Machacek$^3$, C. Jones$^3$, S.W. Randall$^3$}
  \affil{$^1$ Max Planck Institute for Astrophysics\\
    Karl-Schwarzschild-Stra{\ss}e 1, 85748 Garching bei M{\"u}nchen, Germany}
  \affil{$^2$ Department of Physics, Washington University in St. Louis\\
    One Brookings Drive, St. Louis, MO 63130}
  \affil{$^3$ Harvard-Smithsonian Center for Astrophysics \\
       60 Garden Street, Cambridge, MA 02138 USA}
\email{ckreisch@wustl.edu} 

\begin{abstract}
We present an analysis of the complex gas hydrodynamics in the X-ray
luminous galaxy cluster RXJ1347.5-1145 caught in the act of merging
with a subcluster to its southeast using a combined~$186$~ks {\it
  Chandra} exposure,~$2.5$~times greater than previous analyses. The 
primary cluster hosts a sloshing cold front spiral traced by four
surface brightness edges $5 \farcs 85^{+0.04}_{-0.03}$~west,
$7 \farcs 10^{+0.07}_{-0.03}$~southeast, $11 \farcs 5^{+1.3}_{-1.2}$~east, 
and $16 \farcs 7^{+0.3}_{-0.5}$~northeast from the primary central
dominant galaxy, suggesting the merger is in the plane of the sky. We 
measure temperature and density ratios across these edges, confirming 
they are sloshing cold fronts. We observe the eastern edge of the 
subcluster infall shock, confirming the observed subcluster is traveling from 
the southwest to the northeast in a clockwise orbit. We measure a 
shock density contrast of~$1.38^{+0.16}_{-0.15}$~and infer a Mach 
number $1.25\pm0.08$ and a shock velocity of 
$2810^{+210}_{-240}$\,\kms. Temperature and entropy maps 
show cool, low entropy gas trailing the subcluster in a southwestern tail, consistent with core shredding. Simulations suggest a perturber in the plane of the sky on a clockwise orbit would produce 
a sloshing spiral winding counterclockwise, opposite to that observed. The most compelling solution to this discrepancy is that the observed southeastern subcluster is on its first passage, shock heating gas during its clockwise infall, while the main cluster's clockwise cold front spiral formed from earlier encounters with a second perturber orbiting counterclockwise.

\end{abstract}

\keywords{galaxies: clusters: general -- galaxies: interactions --
  X-rays: galaxies: clusters -- galaxies: clusters: individual (RXJ$1347.5-1145$) 
}


\section{INTRODUCTION}
\label{sec:introduction}

The growth of structure across cosmic time provides an important probe
of cosmology. The evolution of the galaxy cluster mass function with
redshift places significant constraints on the dark energy equation of
state, with the greatest sensitivity found at the high mass
end \citep{mass_function1, mass_function2}. 
Since the baryon distribution in massive galaxy  clusters is dominated
by hot, X-ray emitting gas tracing the cluster dark matter potential, 
X-ray observations provide a natural tool to identify clusters. If the 
cluster is relaxed and the gas in hydrostatic equilibrium,
these observational measures allow us to calculate the cluster mass. 
However, in the current hierarchical model for structure formation,
massive clusters grow through mergers between galaxy groups and less massive
clusters, that formed along overdense filaments in the cosmic
web. Thus mergers are expected to be common. Kinetic energy from
the merging partner is converted mainly to thermal 
energy in the form of shocks and turbulence in the cluster gas, 
profoundly affecting the evolution of the ICM 
\citep{Fujitab, Fujitaa, Fujitac, zuhone2010, zuhone2011}. Non-hydostatic gas 
motions induced by the merger (`sloshing') are long lasting on timescales
of order gigayears. Their characteristic signatures in 
X-ray observations, i.e. multiple surface brightness discontinuities
(cold fronts) and/or  sweeping spiral features in temperature and 
surface brightness,  have been noted in many galaxy groups and 
clusters \citep[e.g. see review by][]{markevitch2007, ascasibar2006, paterno-mahler, randall_galaxy, machacek_galaxy}.

The study of these merger-induced gas motions and entropy exchanges 
in massive clusters are key to understanding (1) the properties and
evolution of the ICM and (2) how the 
departures from hydrostatic equilibrium affect uncertainties in the cluster 
mass determinations used to constrain cosmology.

\begin{figure*}[t!]
\begin{center}
\includegraphics[width=\textwidth]{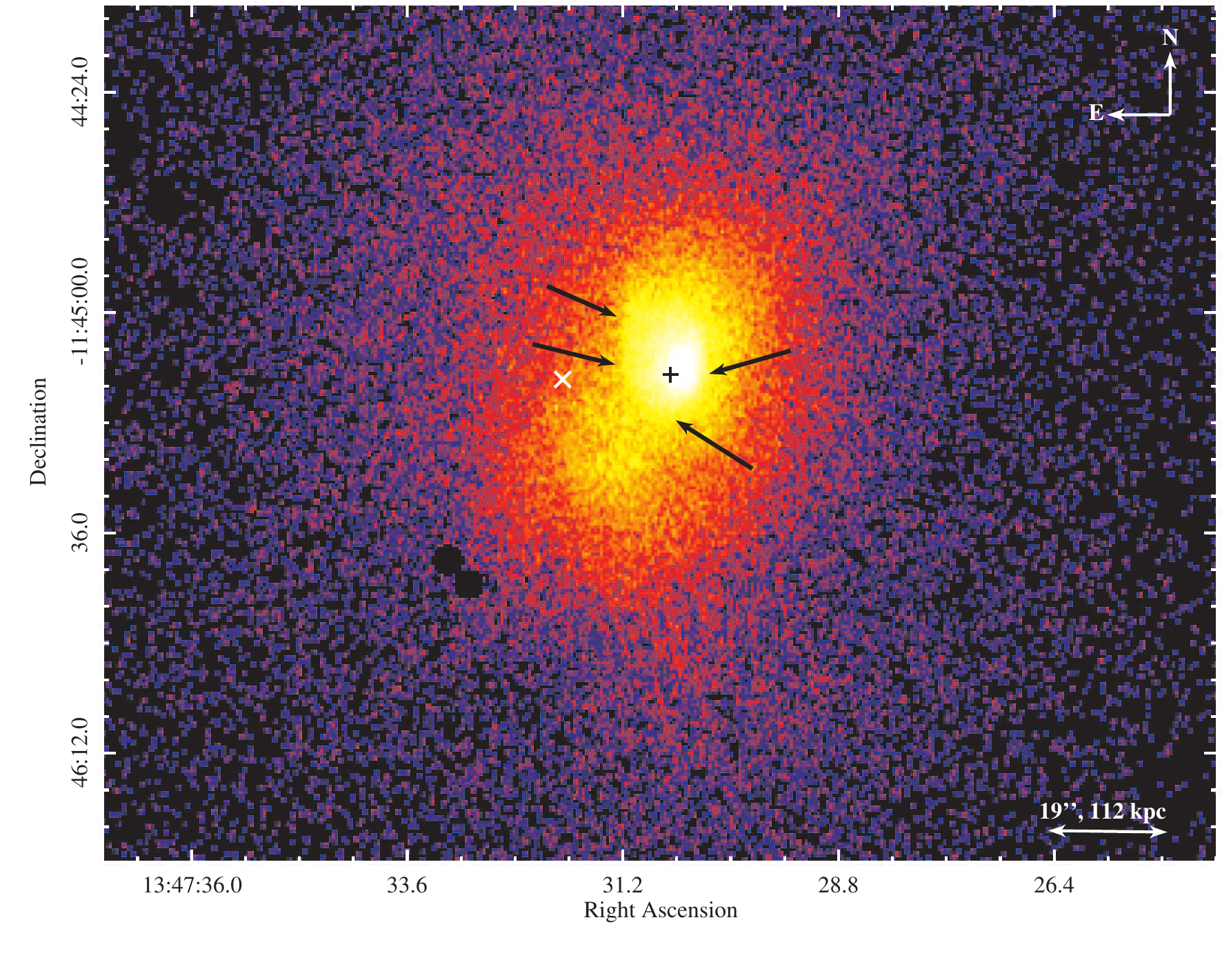}
\caption{Chandra background subtracted, exposure corrected, combined 0.5-2.5 keV full resolution ($1\,\mbox{pixel}=0 \farcs 492\times 0 \farcs 492$) image of RXJ1347 with log scale. The primary cD galaxy is labeled with a black cross and the subcluster cD galaxy is labeled
with a white x. Arrows point to edges forming a spiral pattern.}
\label{fig:xray}
\end{center}
\end{figure*}

Outstanding questions include:

\begin{itemize}
\item{What role do merger shocks play in heating the cluster gas 
  \citep{zuhone2010}?}
\item{ Does sloshing inhibit or promote cooling in the cluster 
core \citep{zuhone2010, zuhone2011}?}
\item{  What can the morphology of the cold fronts and spirals 
tell us about the merger history of the cluster, cluster magnetic fields, or
microphysics of the gas \citep[see, e.g. review by][]{markevitch2007, roedigerb, roedigera, roedigerc, roedigerd, roedigere}}?  
\item{ Does gas sloshing amplify magnetic fields and reaccelerate electrons, 
creating radio mini-halos found to be correlated with cold front edges 
in some clusters \citep{gitti2002, gitti2004, mazzotta2008, zuhone2013}?}
\end{itemize} 
Progress on these questions will be found by a comparison of 
deep X-ray observations of relatively nearby ($z \lesssim 0.5$), massive,  
merging galaxy clusters, where gas temperatures and densities at cold fronts, 
shocks, and sloshing spirals can be well studied, 
with high resolution numerical simulations that connect the
observed X-ray features with the orbital history of the merger  
and the microphysical properties of the surrounding gas. 

In this paper we use new and archival X-ray observations and 
archival Sunyaev-Zel'dovich effect (SZE) 
observations from the literature to characterize
the cold fronts, shocks, and sloshing in RXJ$1347.5-1145$ (RXJ1347). 
 With a  $2 - 10$\,keV X-ray luminosity of $6 \times 10^{45}$\ergs, 
RXJ1347 is the most X-ray luminous galaxy cluster
found in the Rosat All Sky Survey \citep{schindler} and 
is still one of the more X-ray luminous galaxy clusters known. 
RXJ1347 is a massive, highly evolved cool core cluster, yet still 
actively merging with a less massive subcluster. Cool core clusters 
often show signatures of gas sloshing due to the presence of easily 
disturbed low entropy gas in their cores
\citep{markevitch2007}. This makes RXJ1347 an
ideal laboratory to study the on-going 
evolution of the bright end of the cluster luminosity
function. It  has been widely studied across many wavebands 
\citep[see, e.g.][for a review]{johnson2012}.  
Optical spectroscopic surveys of its galaxy population yield a 
velocity dispersion and dynamical mass within $r_{ 200}$ for RXJ1347 
of $\sigma = 1163 \pm 97$\,\kms and 
$1.16^{+0.32}_{-0.27} \times 10^{15}\mbox{$\Ms$}$, respectively 
\citep{lu2010, cohen}. \citet{lu2010} also found that RXJ1347 
resides on a large scale
filament, separated from a less massive galaxy cluster (RXJ1347-SW) by 
$\sim 7$\,Mpc in projection and $4000$\,\kms in radial velocity and 
with an excess of galaxies in between. Although the large velocity
difference between the two clusters makes it unlikely that they are 
interacting, continued merging activity of RXJ1347 with other subclusters
embedded within the same large scale filament would be expected in 
hierarchical cosmological models. The cluster
K-band light is dominated by the two brightest cluster galaxies (BCGs).
One, at (RA $13^h47^m30.7^s$, Dec $-11^{\circ}45'10 \farcs 1$ ) near the peak of the
X-ray surface brightness, is assumed to be the central dominant
galaxy of the primary cluster, and the second, at (RA $13^h47^m31.9^s$, 
Dec $-11^{\circ}45'10 \farcs 9$) lies $18''$  to the east. Given the second
galaxy's relative 
radial velocity difference of only $\sim 100$\kms with respect to the primary
cluster BCG, it has been interpreted as the central dominant galaxy 
of a merging subcluster. Surface mass density maps from
weak and strong lensing show a general elongation in the mass
concentration in the direction of the second BCG. However, they do not show a 
clear mass concentration peak there \citep{bradac2008, schmidt}. 

Early X-ray observations by {\it Chandra} \citep{allen} 
and {\it XMM-Newton} \citep{gitti2004} found excess X-ray 
emission~$\sim20''$~southeast of the primary BCG, which we now 
interpret as emission from hot gas in a merging subcluster. 
Deeper {\it Chandra} observations have revealed two sloshing cold 
fronts in the primary cluster as a result of a recent 
merger \citep{johnson2012}.

Sunyaev-Zel'dovich observations have played a critical role in 
understanding the merging subcluster's gas dynamics. Observations 
by \citet{komatsu2001} with the Nobeyama 45 m telescope at~150~GHz 
first showed a significant SZ decrement~$\sim 20''$~southeast of the 
primary cluster, suggesting RXJ1347 is a disturbed system that has 
recently undergone a merger \citep{kitayama2004}. Higher resolution 
observations with the MUSTANG camera on the Green Bank Telescope 
at~90~GHz indicated the SZ decrement likely corresponds to shock 
heated gas from the merging subcluster 
\citep{mason2010, korngut2011}. 
A higher temperature and characteristic high pressure 
gradient support the interpretation that the region between the
clusters is a shock front \citep{mason2010}. RXJ1347 also hosts a
radio mini-halo centered on its primary cluster. While gas sloshing 
causes electron re-acceleration in the mini-halo, \citep{mazzotta2008}, 
\citet{ferrari2011} used observations by the Giant Metrewave Radio 
Telescope at 237 MHz and 614 MHz to find that excess radio emission 
in the southeast section of the mini-halo is likely caused by a 
propagating shock front corresponding to the shock-heated gas 
and SZ decrement previously found. 

At a redshift of $z=0.451$, RXJ1347 is close enough that the disturbed 
X-ray morphology resulting from recent mergers can be observed in
detail with the high angular resolution of the {\it Chandra} X-ray 
Observatory. At this high angular resolution, our combined $186$\,ks 
{\it Chandra} exposure ( $> 2.5$ times
deeper than previous {\it Chandra} observations) allows 
us to identify features not seen before, 
to better constrain the ICM gas hydrodynamics and, by comparison with 
simulations, RXJ1347's  merger history. In \S\ref{sec:obs} we discuss 
our reduction of the observational data and general analysis 
techniques.  In \S\ref{sec:qual} we analyze the mean and asymmetric 
gas properties in the primary cluster. We discuss the properties of 
the merging subcluster in \S\ref{sec:subcluster}, and we compare our 
results to simulations in \S\ref{sec:sim} to discern the cluster 
merging history. 

For the standard $\Lambda$
dominated cold dark matter cosmology and assuming 
$H_0 = 70$\kmsmpc, $\Omega_m = 0.3$, $\Omega_\Lambda = 0.7$,
the redshift $z=0.451$ for
RXJ1347.5-1145 corresponds to a luminosity distance of $2504$\,Mpc 
and angular scale $1'' = 5.77$\,kpc \citep{cosmocalc}. All WCS coordinates 
are J2000, and uncertainties are at $90 \%$ CL unless otherwise specified.


\section{OBSERVATIONS AND DATA REDUCTION}
\label{sec:obs}

\begin{deluxetable}{ccccc}
\tablewidth{0pc}
\tablecaption{{\it Chandra} X-ray Observations \label{tab:obs}}
 \tablehead{ \colhead{ObsID} & \colhead{Exposure Time} & \colhead{Bkg Norm}& \colhead{Date} & \colhead{PI} \\
     & (ks) }
 \startdata
  \dataset [ADS/Sa.CXO#obs/3592] {3592} & $49.643$ &$0.870$ & $2003-09-03$ 
& L. van Speybroeck \\
  \dataset [ADS/Sa.CXO#obs/14407] {14407} & $56.076$ &$0.769$ &$2012-03-16$ 
& C. Jones \\
  \dataset [ADS/Sa.CXO#obs/13999] {13999} & $47.817$ &$0.842$ &$2012-05-14$ 
& C. Jones \\
  \dataset [ADS/Sa.CXO#obs/13516] {13516} & $32.629$ &$0.772$ &$2012-12-11$ 
& S. Murray
  \enddata
 \tablecomments{{\it Chandra} observations used in this analysis. All
   observations were taken in VFAINT mode with ACIS-I at
   aim point. Column $2$ lists the effective exposure after
   excluding periods of anomalously high and low count rates 
   (see \S\protect\ref{sec:obs}). Column $3$ lists the additional 
   normalization needed for blank sky backgrounds to match the
   observation count rate in the $10-12$\,keV energy band.}
\end{deluxetable}

In Table \ref{tab:obs} we list the data sets used in this analysis.  
All observations were taken by the Advanced CCD 
Imaging Spectrometer \citep[ACIS][]{acisa,acisb} on board 
the {\it Chandra} X-ray 
Observatory with the ACIS-I array at aim point.  Using the {\it
Chandra} Interactive Analysis of Observations suite of 
analysis tools \citep[CIAO 4.6][]{ciao} and CALDB 4.6.1.1,  
the data were reprocessed with the latest calibrations including 
updated gains and
observation dependent bad pixel files, and corrections for the 
charge transfer inefficiency on the ACIS CCDs, the time-dependent 
build-up of contaminant on the optical filter, and the secular drift of 
the average pulse-height amplitude for photons of fixed energy
(tgain).  The data were then filtered to reject bad array patterns 
(grades $1$, $5$, and $7$), and to remove 
data flagged by the VFAINT mode as having excessive counts in 
border pixels surrounding event islands. Particle flares were cleaned 
from the data with the tool {\em lc\_clean}, using a $3\sigma$ 
clipping algorithm to calculate the mean count rate in the
$0.3-12$\,keV energy band, and then rejecting 
time periods in which the count rate fell more than $20\%$ above or below 
the mean. These effective exposure times for each observation 
are listed in Table \ref{tab:obs}, resulting in a combined useful
exposure for our analysis of $186,165$\,s. 

Source free background sets, provided by the Chandra X-ray Center, 
appropriate for the ACIS-I detector and date of observation, were 
reprojected onto each observation, and renormalized to match the 
observation count rate in the $10 - 12$\,keV energy band where particle 
backgrounds dominate. These additional normalization factors are also 
listed in Table \ref{tab:obs}.  Background subtracted, exposure corrected
mosaiced X-ray surface brightness images were then created in various 
energy bands using the CIAO script {\em flux\_image} and other CIAO tools. 

We used the broad band ($0.5- 7$\,keV) mosaiced X-ray surface
brightness image to identify `by eye' $77$ point sources in the 
combined field of view. Since we are interested in the properties of
the diffuse emission from the cluster gas, these sources
were excised from the data. We compared our point source list to that
given in the Chandra Source Catalog \citep{evans} for this
region, and found that our removal of point sources was more conservative. 
We also excluded from our analysis a linear artifact (read-out streak), 
seen in the mosaiced image to extend from
the bright X-ray peak at the cluster center to the northeast and
southwest (at $110^\circ$ and $290^\circ$, respectively, measured
counter-clockwise from west). Since we are
interested in the properties of the diffuse emission in the cluster
rather than the central bright source, excluding this artifact 
will not affect our results.  
 
\begin{figure}[tp!]
\begin{center}
\includegraphics[width=0.475\textwidth]{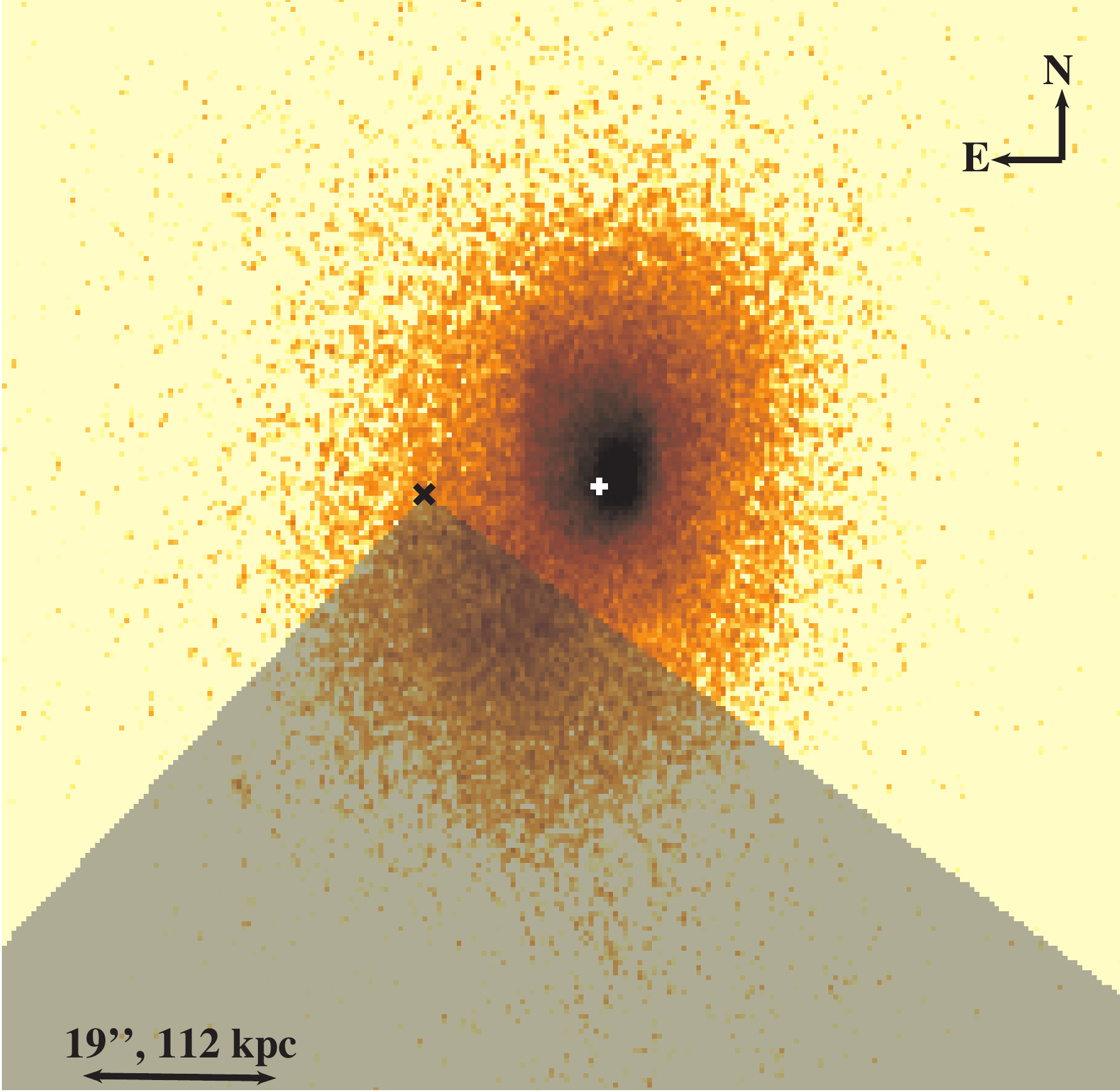}
\caption{We show the flux image from Figure \ref{fig:xray} with a mask
  covering the region excluded from the mean primary cluster gas 
distribution fit. The figure has a log scale with the primary
(subcluster) cD galaxy labeled as a white cross (black X).} 
\label{fig:mask_resid}
\end{center}
\end{figure}

\section{Primary Cluster}
\label{sec:qual}
The $186$~ks~{\it Chandra} background subtracted, exposure corrected, combined
$0.5-2.5$~keV full resolution X-ray surface brightness image is
displayed in Figure \ref{fig:xray}. We used the~$0.5-2.5$~keV energy
band for imaging analysis, even though the cluster is hot, because of
lower signal to noise in the higher ($2.5-7.0$~keV) energy bands. The asymmetries
in the surface brightness suggest the presence of dynamic, complex gas 
motions in both the
primary cluster and merging subcluster. The subcluster is visible 
southeast of the primary cD galaxy, with its X-ray peak
$\sim 22''$ distant from the primary cluster BCG, as seen in previous
X-ray and SZ observations \citep[see, e.g.][]{johnson2012,
  komatsu2001, kitayama2004, ota}. In this section we focus on
the properties of the primary cluster gas, deferring the discussion
of the subcluster to Section \ref{sec:subcluster}. 
Edges in the surface brightness trace a spiral feature
in the primary cluster. We have noted these edges, as well as the
centers of the primary and subcluster cD galaxies in
Figure \ref{fig:xray}. To understand these gas 
asymmetries in the primary cluster, we begin with an analysis of its
mean gas distribution. 

\subsection{Gas Distribution}
\label{sec:mod}

We use {\it Sherpa} to fit a $2$ dimensional elliptical beta model of surface
brightness to the RXJ$1347$ $0.5-2.5$ keV full resolution ($1$
pixel$=0 \farcs 492$) counts image:
\begin{eqnarray}
\label{eq:ell_beta}
S_x\left(r\right) = A\left[1+\left(\frac{r}{r_0}\right)^2\right]^{-\alpha}
\end{eqnarray}
with
\begin{eqnarray}
r\left(x,y\right) = \frac{\sqrt{x_{new}^2(1-\epsilon)^2+y_{new}^2}}{1-\epsilon}
\end{eqnarray}
\begin{eqnarray}
x_{new} = \left(x-x_0\right)\cos\left(\theta \right)+\left(y-y_0\right)\sin\left(\theta \right)
\end{eqnarray}
\begin{eqnarray}
y_{new} = \left(y-y_0\right)\cos\left(\theta \right)-\left(x-x_0\right)\sin\left(\theta \right)
\end{eqnarray}
and where $x_0$ and $y_0$ are the coordinates of the X-ray
peak, A is the amplitude at $x_0$ and $y_0$, $\epsilon$ is
ellipticity, $\theta$ is the angle of
ellipticity, $r_0$ is the core radius, and $\alpha$ is the power law
index. An exposure map for the full resolution $0.5-2.5$
keV image was included in the fit. 

To obtain parameters descriptive of only the primary
cluster rather than the system with both clusters, a sector covering the
subcluster centered on the second brightest cluster galaxy 
(RA $13^h47^m31.9^s$, Dec $-11^{\circ}45'10\farcs 9$) spanning an angle 
of $227^{\circ}$ to $324^{\circ}$ counterclockwise from west was
excluded from the image using CIAO tools (see
Figure \ref{fig:mask_resid}). The sector's size was chosen given the
cone-like appearance of the subcluster in Figure \ref{fig:xray}. We
choose to exclude the subcluster's gas contribution rather than
jointly fit both clusters' distributions because there is little
diffuse subcluster gas contribution to the primary cluster outside the 
cone surrounding the subcluster. Parameters for modeling the masked image were found by first using \textit{Sherpa}'s Monte Carlo optimization method with the Cash statistic to roughly locate the global minimum \citep{cash, montecarlo}. The Levenberg-Marquardt minimization technique was then used with the Monte Carlo best-fit as starting points to precisely locate the minimum \citep{levmar}. Core radius $r_0$, $x_0$ and $y_0$ positions, ellipticity $\epsilon$, angle of ellipticity $\theta$, amplitude $A$, and model power law index $\alpha$ were allowed to vary simultaneously during each minimization. Parameters for the best fit are listed in Table \ref{tab:param}. 

\begin{deluxetable}{cccc}
 \tablewidth{0pc}
 \tablecaption{2 Dimensional Elliptical Beta Model Parameters: Primary Cluster\label{tab:param}}
 \tablehead{\colhead{Parameter} & \colhead{Best-Fit} & \colhead{Lower
     Bound} & \colhead{Upper Bound} 
      }
 \startdata
  $r_0$  & $5.1$ & $-0.1$ & $0.04$ \\
  A      & $1.6\times10^{-6}$ & $-1.6\times10^{-8}$ & $4.5\times10^{-8}$ \\
  $\alpha$ & $1.12$ & $-0.005$ & $0.004$ \\
  $x_0$  & $13:47:30.7$ & $-0.01$ & $0.1$ \\
  $y_0$  & $-11:45:08.1$ & $-0.03$ & $0.1$ \\
  $\epsilon$ & $0.2$ & $-0.005$ & $0.006$ \\
  $\theta$ & $88$ & $-1$ & $1$ 
 \enddata
 \tablecomments{Best-fit {\it Sherpa} parameters with uncertainties
   for the primary cluster fit. $x_0$ and $y_0$
   are given in RA, Dec. Core radius 
   and its uncertainties, as well as $x_0$ and $y_0$ uncertainties,
   are in arcsec, while amplitude (A) is in 
   photons pix$^{-1}$~cm$^{-2}$~s$^{-1}$. $\alpha$ and $\epsilon$
   are unitless, and $\theta$ has units of degrees. Uncertainties for the fit are $90 \%$ CL.    
 }
 \end{deluxetable}

We can find the value of the familiar $\beta$ parameter by equating $\alpha=3\beta-1/2$ \citep{beta}. Given the best-fit $\alpha$ value of $1.12$, $\beta=0.54$ for a core radius of $r_0=29.4^{+0.2}_{-0.6}$~kpc. The average gas density can be modeled using our $\beta$
value:
\begin{eqnarray}
n_{gas}\left(r\right) = n_{gas}\left(0\right)\left[1+\left(\frac{r}{r_0}\right)^2\right]^{-3\beta/2}
\end{eqnarray}

Our mean $\beta$-model fit agrees well with that found by \citet{allen} ($\beta = 0.535 \pm 0.003$, $r_0 = 29.2 \pm 0.7$\,kpc), from fitting the $0.3-7$\,keV image excluding the southeast subcluster quadrant. Previous shallower 
observations by \citet{allen} and \citet{gitti2007b} found that
the cluster appears, on average, symmetric and relaxed. Our model shows a small elongation in the N-S direction, consistent with qualitative features seen in the deep X-ray image in Figure \ref{fig:xray}.

\begin{figure*}[tp!]
\begin{center}
\includegraphics[width=0.500\textwidth]{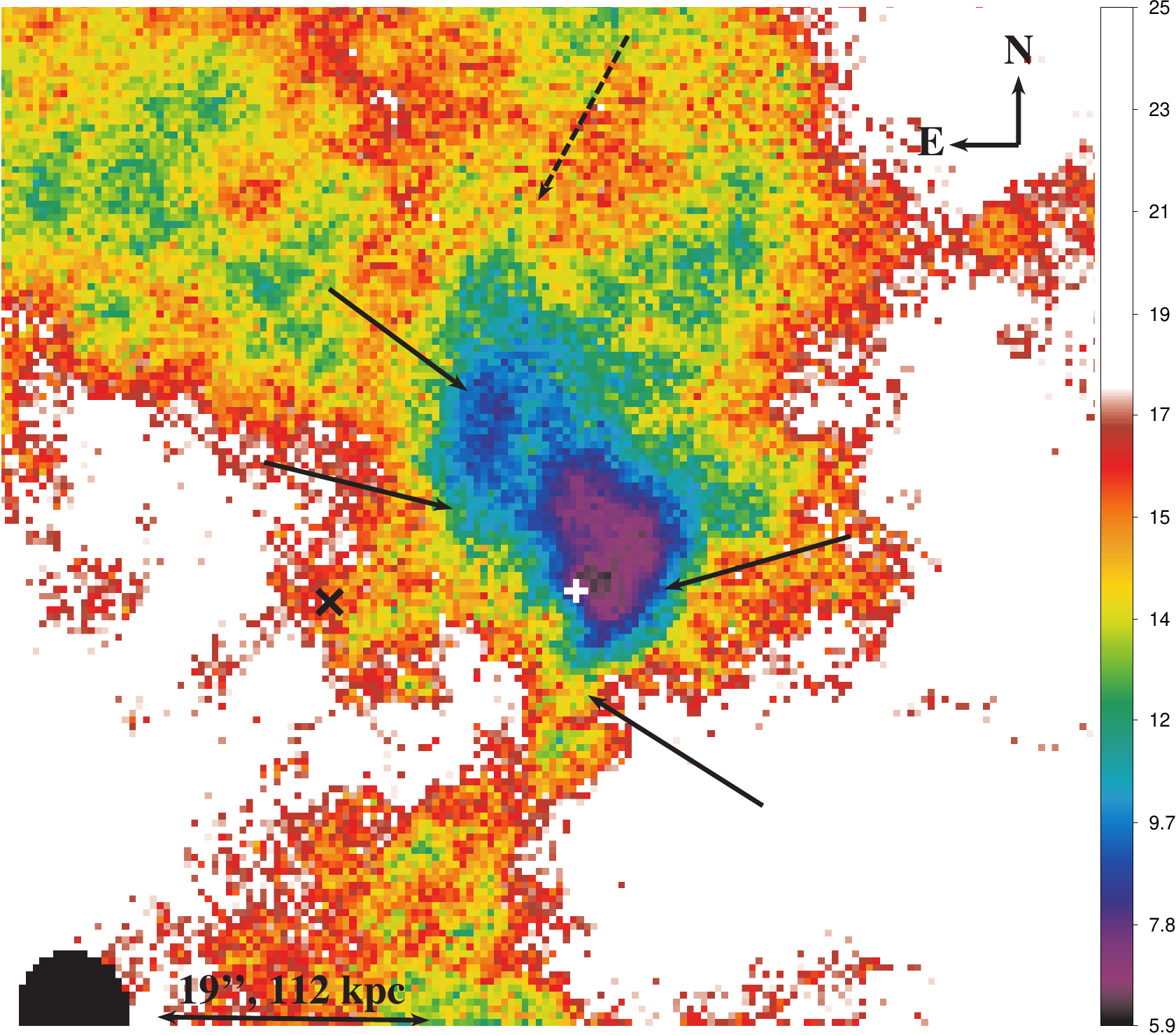}
\includegraphics[width=0.450\textwidth]{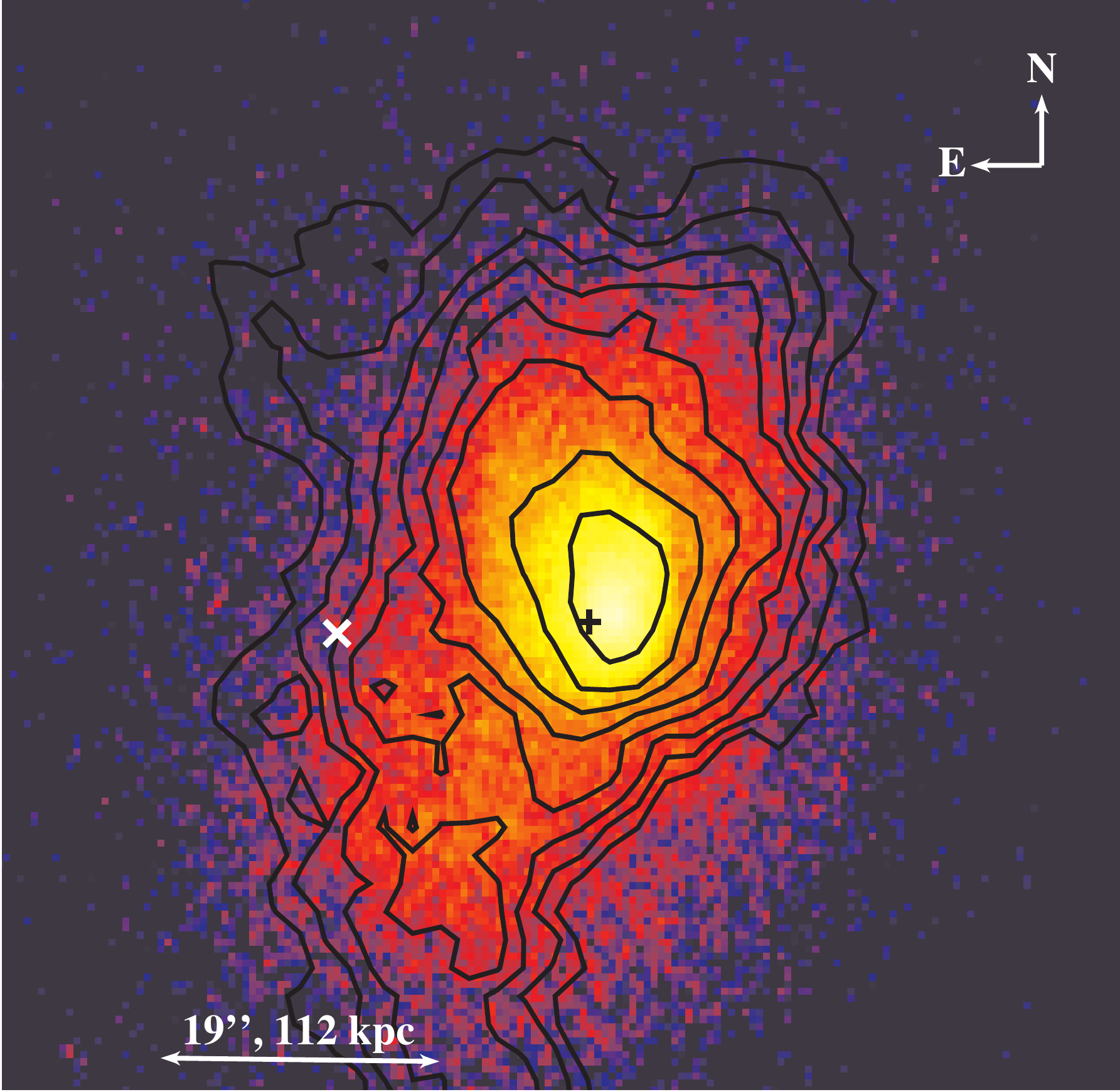}
\caption{(Left) Temperature map of RXJ1347. The rainbow color scale
  increases in temperature from black (lowest) to white (highest) in keV. 
The black circle is a removed point source. Arrows point to the cold 
fronts labeled in Figure \ref{fig:xray}. The fan region is shown by a dashed arrow. $1\,\sigma$ uncertainties in the fits range from $8\%$ for temperatures $\sim6.5$ keV in the core to $25-30\%$ for temperatures $\gtrsim 20$ keV in the cluster outskirts. (Right) Entropy contours overlaid on the full resolution combined flux image from Figure \ref{fig:xray} matched in WCS
coordinates to the temperature map. Low entropy contours trace the 
cold fronts and fan that form part of the spiral. The +(X) symbol 
denotes the primary cluster (subcluster) BCG, respectively, in both panels.}
\label{fig:cfpos}
\end{center}
\end{figure*}

\begin{deluxetable*}{cccccc}
 \tablewidth{0pc}
 \tablecaption{Primary Cluster Cold Front Profile Regions\label{tab:ellreg}}
 \tablehead{\colhead{Region} & \colhead{Semi-Minor Axis} & \colhead{Semi-Major Axis} & \colhead{Position Angle} & \colhead{Angle A} &
   \colhead{Angle B} \\
      & $\left( \mbox{arcsec}\right)$ & $\left( \mbox{arcsec}\right)$ & $\left( \mbox{deg}\right)$ & $\left( \mbox{deg}\right)$ & $\left( \mbox{deg}\right)$ }
 \startdata
  West &
  $5.1$ & $8.5$ & $74$ & $315$ & $42$\\
  Southeast &
  $6.2$ & $8.4$ & $137$ & $183$ & $275$ \\ 
  East &
  $6.8$ & $16.3$ & $107$ & $127$ & $152$ \\
  Northeast &
  $6.6$ & $24.0$ & $99$ & $109$ & $127$
 \enddata
 \tablecomments{Bounding ellipses trace sector edges. All sectors
   were centered at primary cD RA $13:47:30.7$, Dec $-11:45:10.1$ and 
   subtend the angle from A to B, moving counter-clockwise from west. 
   Position angles are defined as the angle between the semi-major
   axis and west.
 }
 \end{deluxetable*}

\begin{deluxetable*}{ccccccc}
 \tablewidth{0pc}
 \tablecaption{Cold Front Broken Power Model Parameters\label{tab:cfparam}}
 \tablehead{\colhead{Region} & \colhead{Inner Slope} &
   \colhead{Outer Slope} & \colhead{Location of Edge} &
   \colhead{$\frac{\rho_2}{\rho_1}$} & \colhead{$\frac{T_2}{T_1}$} &
   \colhead{$\frac{P_2}{P_1}$} \\
      & &  & $\left( \mbox{arcsec}\right)$ & }
 \startdata
  West & $-0.65^{+0.10}_{-0.08}$ & $-1.14^{+0.05}_{-0.04}$ & $5.85^{+0.04}_{-0.03}$ & $1.80^{+0.10}_{-0.08}$ & $0.56^{+0.16}_{-0.13}$ & $1.00^{+0.29}_{-0.24}$ \\
  Southeast & $-0.78 \pm 0.09$ & $-0.72^{+0.06}_{-0.05}$ & $7.10^{+0.07}_{-0.03}$ & $1.62^{+0.15}_{-0.13}$ & $0.65^{+0.29}_{-0.22}$ & $1.06^{+0.48}_{-0.37}$ \\
  East & $-0.53 \pm 0.08$ & $-1.36^{+0.06}_{-0.14}$ & $11.5^{+1.3}_{-1.2}$ & $1.51^{+0.11}_{-0.21}$ & $0.66^{+0.48}_{-0.24}$ & $0.99^{+0.73}_{-0.39}$ \\
  Northeast & $-0.76 \pm 0.04$ & $-1.57 \pm 0.06$ & $16.7^{+0.3}_{-0.5}$ & $1.44 \pm 0.09$ & $0.58^{+0.27}_{-0.16}$ & $0.83^{+0.39}_{-0.24}$
 \enddata
 \tablecomments{Parameters correspond to the best-fit density model
   for the elliptical profiles. $\rho_1$, $T_1$, and $P_1$ refer to the density,
   temperature, and pressure outside the cold front, and $\rho_2$,
   $T_2$, and $P_2$ represent the density,temperature, and pressure
   inside the cold front. Uncertainties are presented at $90 \%$
   confidence. Temperatures were fit on each side of the edges 
  (see Table \protect\ref{tab:cftemp}). 
 }
 \end{deluxetable*}

\begin{deluxetable}{ccc}
 \tablewidth{0pc}
 \tablecaption{Spectral Local Background Regions\label{tab:lbgreg}}
 \tablehead{\colhead{ObsID} & \colhead{$x_0$} &
   \colhead{$y_0$} \\
      & RA & dec }
 \startdata
  $13516$ & $13:47:41.1$ & $-11:44:05.2$ \\
  $3592$ & $13:47:41.1$ & $-11:44:05.2$ \\
  $13999$ & $13:47:40.4$ & $-11:44:46.9$ \\
  $14407$ & $13:47:39.1$ & $-11:43:59.3$
 \enddata
 \tablecomments{Unrotated rectangles with sides of $39 \farcs 9$ and
   $38 \farcs 9$ were used as local backgrounds when extracting
   spectra from regions. }
 \end{deluxetable}

\subsection{Characterization of Gas Asymmetries}
\label{sec:asym}

Displacement of the X-ray surface brightness peak to the west of the 
primary cluster cD galaxy indicates the core has most recently sloshed 
to the west (see Figure \ref{fig:xray}). We show a projected temperature 
map in the left panel of Figure \ref{fig:cfpos}. The temperature map was created by growing regions around each pixel in the $0.6-9$ keV mosaiced image until the region contained at least 3000 counts. The spectrum was then fit with an absorbed APEC model with Galactic absorption and \citet{anders} abundance fixed at $0.5\,\mbox{Z}_{\odot}$ \citep{ota}. $1\,\sigma$ uncertainties in the fits range from $8\%$ for temperatures $\sim6.5$ keV in the core to $25-30\%$ for temperatures $\gtrsim 20$ keV in the cluster outskirts.

The temperature map in Figure \ref{fig:cfpos} shows complicated structures. A temperature dip 
west of the primary BCG indicates the sloshing core's presence. High
temperatures in the southwest and southeast at large radii (the white
regions in Figure \ref{fig:cfpos}) relative to 
temperatures at similar radii to the north and northeast could
be due to shock heating from a previous encounter.

Wrapping around the primary BCG in the X-ray flux image is a clockwise 
spiral that is elongated to the north. We denote these spiral edges in 
Figure \ref{fig:xray}. Both the west and east cold fronts
identified by \citet{johnson2012} form part of the spiral. Edges in 
the projected temperature image align well with those in the X-ray 
flux image (see arrows in Figures \ref{fig:xray} and \ref{fig:cfpos}). 
In the right panel of Figure \ref{fig:cfpos}, we overlay pseudo-entropy 
($s=TS_X^{-1/3}$) contours on the flux image. The contours show low 
entropy gas in the primary cluster core that traces the cold fronts. 
A diffuse, cool fan extends from the northeastern spiral elongation 
in both the temperature and flux images. The low entropy contours 
also elongate along this fan feature. However, the temperature map is coarse, giving only a qualitative picture of the complicated temperature structure in the cluster and serving as a guide for more careful spectral modeling of regions of interest.

\subsection{Cold Fronts}
\label{sec:coldfront}

Elliptical surface brightness profiles of the combined, background
subtracted, exposure corrected $0.5-2.5$ keV full resolution flux image
were taken to characterize edge properties. The profiles were taken in 
four sectors, each centered on the primary cluster BCG, based on
features seen in the X-ray and temperature images described 
in \S\ref{sec:asym}. The ellipses used to construct the profiles 
were chosen congruent to a bounding ellipse that traces the edge in 
each sector. See Table \ref{tab:ellreg} for parameters defining each 
sector and the bounding ellipses used to trace the edges in each sector. 

We integrate a broken power law model of density along the line of
sight to fit the elliptical surface brightness profiles across each
edge. The density model is given by:
\begin{align}
\label{eq:cf1}
 n_{gas}\left(r\right)&=n_{e,0}\frac{\rho_2}{\rho_1}\left(\frac{r}{r_s}\right)^{\nu_2} \qquad &\mbox{for}\; r<r_s \\
\label{eq:cf2}
n_{gas}\left(r\right)&=n_{e,0}\left(\frac{r}{r_s}\right)^{\nu_1} \qquad &\mbox{for}\; r>r_s
\end{align}
where $n_{e,0}$ is the overall normalization, $\frac{\rho_2}{\rho_1}$ 
is the density ratio between the inner and outer power law slopes, 
$r_s$ is the location of the edge, and $\nu_2$ and
$\nu_1$ are the inner and outer power law slopes, respectively. 
Results of our fits are summarized in Table \ref{tab:cfparam}.

\begin{deluxetable*}{cccccc}
 \tablewidth{0pc}
 \tablecaption{Cold Front Spectral Regions\label{tab:specreg}}
 \tablehead{\colhead{Sector} & \colhead{Semi-Minor Axis Out} &
   \colhead{Semi-Major Axis Out} & \colhead{Semi-Minor Axis In} &
   \colhead{Semi-Major Axis In} \\
     & $\left( \mbox{arcsec} \right)$ & $\left( \mbox{arcsec} \right)$ & $\left( \mbox{arcsec} \right)$ & $\left( \mbox{arcsec} \right)$ }
 \startdata
  W3 &
  $3.50$ & $5.81$ & $2.70$ & $4.49$ \\
  W2 &
  $5.14$ & $8.53$ & $3.50$ & $5.81$ \\
  W1 &
  $8.00$ & $13.38$ & $5.14$ & $8.53$ \\
  W0 &
  $11.00$ & $18.25$ & $8.00$ & $13.38$ \\
  SE3 &
  $3.68$ & $5.01$ & $1.88$ & $2.56$ \\
  SE2 &
  $6.60$ & $8.99$ & $3.68$ & $5.01$ \\ 
  SE1 &
  $9.50$ & $12.94$ & $6.60$ & $8.99$ \\
  SE0 &
  $12.50$ & $17.03$ & $9.50$ & $12.94$ \\
  E3 &
  $3.69$ & $8.82$ & $0$ & $0$ \\
  E2 &
  $6.82$ & $16.32$ & $3.69$ & $8.82$ \\
  E1 &
  $12.3$ & $29.41$ & $6.82$ & $16.32$  \\
  E0 &
  $19.68$ & $47.06$ & $12.3$ & $29.41$ \\
  NE3 &
  $3.48$ & $12.65$ & $1.72$ & $6.26$ \\
  NE2 &
  $6.61$ & $24.04$ & $3.48$ & $12.65$  \\
  NE1 &
  $14.76$ & $53.68$ & $6.61$ & $24.04$
 \enddata
 \tablecomments{Elliptical segments are restricted to lie within each 
sector and are congruent to the bounding ellipse that traces the 
edge within the sector (see Table \protect\ref{tab:ellreg}). 
 }
 \end{deluxetable*}

\begin{figure}[t!]
\begin{center}
\includegraphics[width=0.5\textwidth]{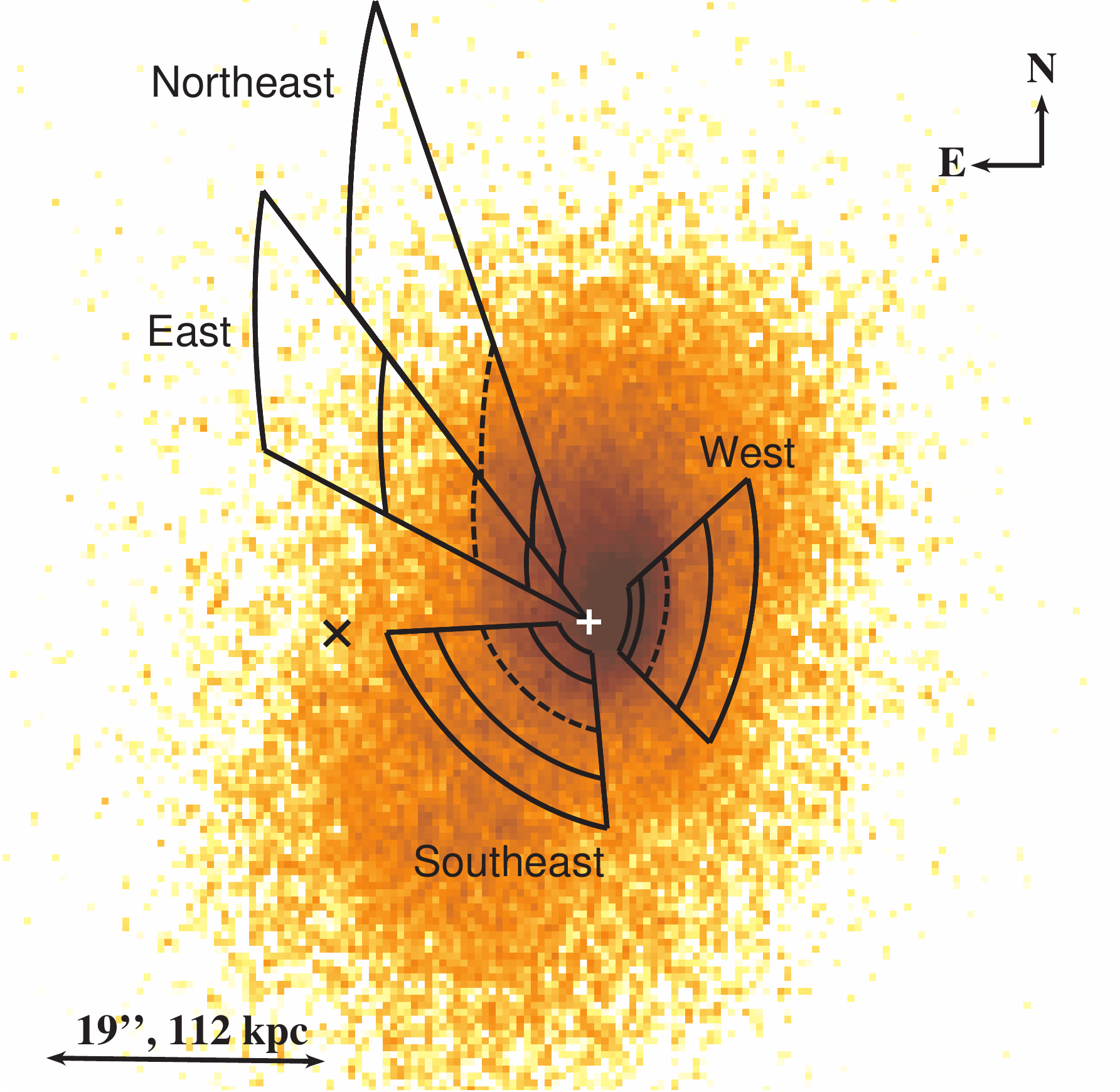}
\caption{Sectors used to extract the elliptical surface brightness
  profiles plotted in the top panel of Figure \protect\ref{fig:eastcf} overlaid
  on the full resolution flux image of RXJ1347 from
  Figure \protect\ref{fig:xray}. Regions used for spectral analysis are shown as 
elliptical arcs concentric to the bounding ellipses (long dashed
lines) listed in Table \protect\ref{tab:ellreg} that trace the observed 
surface brightness edges within each sector.}
\label{fig:entropy_cont}
\end{center}
\end{figure}

\begin{deluxetable*}{ccccccc}
 \tablewidth{0pc}
 \tablecaption{Cold Front Temperatures\label{tab:cftemp}}
 \tablehead{\colhead{Sector}
   & \colhead{Temperature} & \colhead{$\chi^2/\left(\mbox{d.o.f.}\right)$}\\
     & $\left(\mbox{keV}\right)$ & }
 \startdata
 W3$^{\ast}$ & $5.9^{+1.0}_{-0.8}$ & $75.75/84$ \\
 W2 & $7.4^{+1.0}_{-0.9}$ & $204.80/143$ \\
 W1 & $13.3_{-2.7}^{+3.3}$ & $166.91/145$ \\
 W0 & $13.6\pm2.7$ & $130.17/135$ \\
 SE3$^{\ast}$ & $12.8_{-3.4}^{+4.9}$ & $81.48/84$ \\
 SE2 & $13.8_{-3.0}^{+3.6}$ & $169.14/156$ \\
 SE1 & $21^{+8}_{-5}$ & $110.78/130$ \\
 SE0 & $20^{+7}_{-4}$ & $173.42/151$ \\
 E3$^{\dagger}$ & $11.9^{+5.8}_{-3.2}$ & $62.31/69$ \\
 E2$^{\ast}$ & $10.7^{+3.8}_{-2.2}$ & $82.01/83$ \\
 E1$^{\ast}$ & $16.2^{+10.3}_{-4.7}$ & $82.22/77$ \\
 E0$^{\dagger}$ & $9.7^{+4.4}_{-2.3}$ & $51.67/67$ \\
 NE3$^{\dagger}$ & $9.0^{+3.5}_{-1.9}$ & $80.16/66$ \\
 NE2$^{\star}$ & $8.9^{+2.4}_{-1.5}$ & $98.93/100$ \\
 NE1$^{\star}$ & $15.4^{+5.8}_{-3.5}$ & $97.56/103$
 \enddata
 \tablecomments{Temperatures are presented with $90 \%$~CL after being
   fit with fixed Galactic absorption and abundance at 
$0.5\,\mbox{Z}_{\odot}$. Labels of W, SE, E, and NE correspond to 
temperatures in the West, Southeast, East, and Northeast 
profiles. Numerical labels of 3, 2, 1, and 0 correspond to the most 
inward sector, the sector just inside of the cold front, the sector
just outside the cold front, and the sector farthest from the primary 
BCG. We use the following symbols to denote the combined number of 
net counts: $\dagger$: >1000 counts, $\ast$: >1500 counts, 
and $\star$: >2000 counts. Unlabeled sectors have more than 
2500 net counts.  }
 \end{deluxetable*}

We extracted spectra for regions of interest using CIAO script  
{\em  specextract} and chose  $39\farcs 9 \times 38\farcs 9 $ 
rectangular source free regions on the ACIS-I array as local backgrounds for 
each observation (see Table \ref{tab:lbgreg}). Spectra from each of the
   four observations were fit together, but extracted separately for
   each region. Except where noted, we
required a minimum of $2000$ net counts total to constrain the
expected high gas temperatures in this cluster. Spectra of diffuse 
gas were modeled in XSpec 12.8.0, using absorbed APEC models \citep{apec} 
with fixed Galactic absorption ${\rm n_H}= 4.75 \times 10^{20}$\cms
\citep[Leiden/Argentine/Bonn collaboration,][]{snowden} and abundance 
at $0.5\,\mbox{Z}_{\odot}$ \citep{ota}. See Tables \ref{tab:lbgreg} and \ref{tab:specreg} for cold front spectral region 
parameters and Table \ref{tab:cftemp} for temperature values. Spectral 
regions are also shown within sectors in Figure \ref{fig:entropy_cont}.

\begin{figure*}[tp!]
\begin{center}
\includegraphics[width=0.497\textwidth]{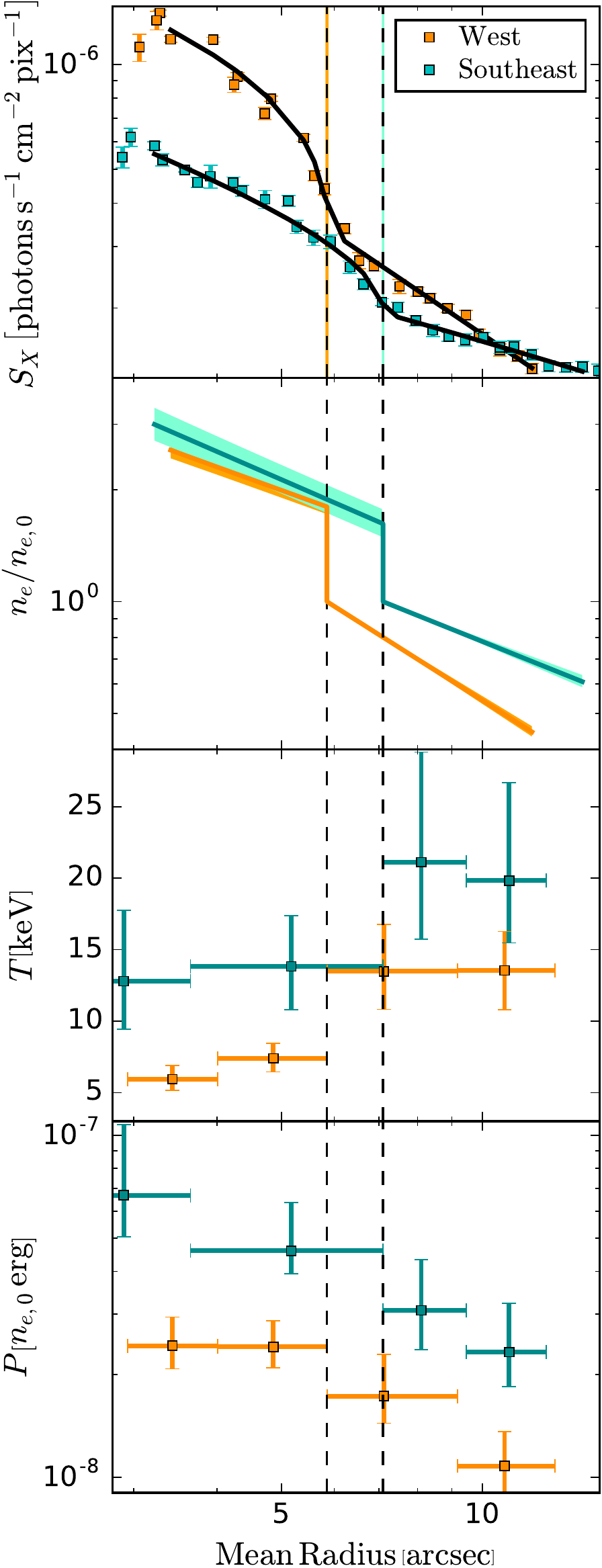}
\includegraphics[width=0.497\textwidth]{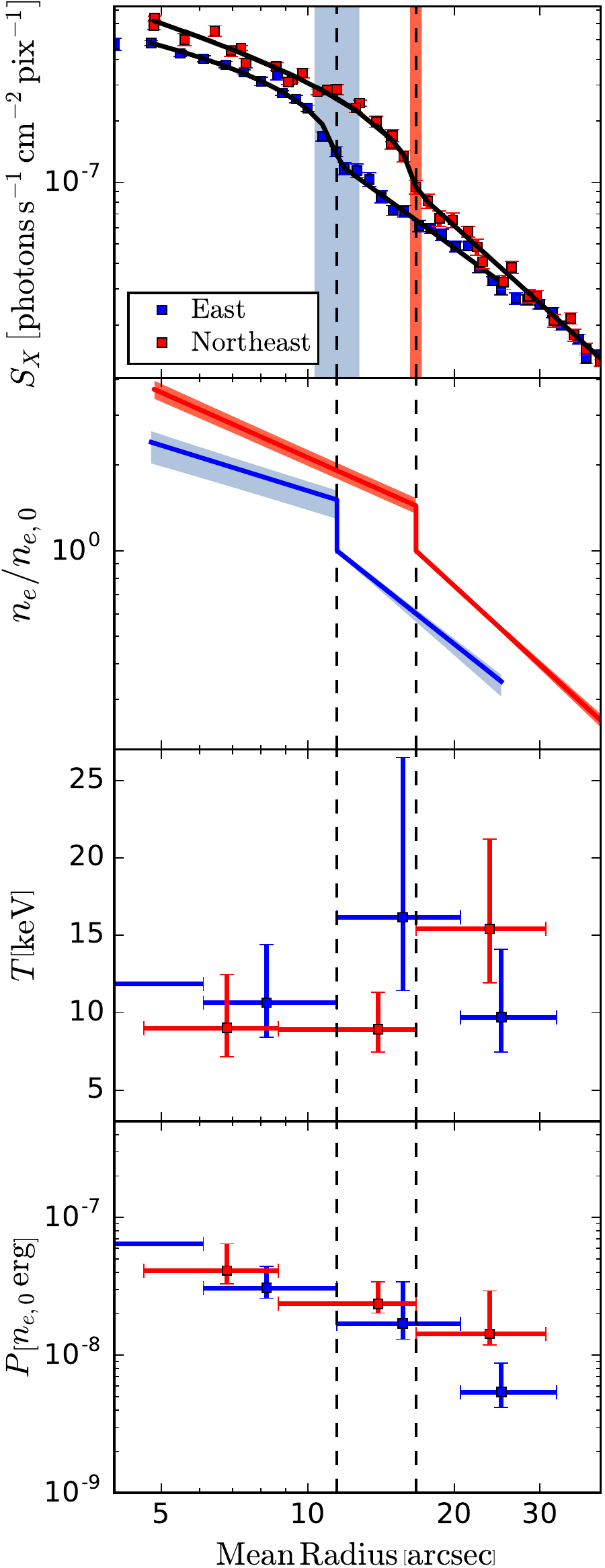}
\caption{Surface brightness profiles, best-fit density profiles,
  temperature profiles, and pressure profiles are displayed for the 4
  identified cold fronts. The left panel shows the west and
  southeast fronts, while the right panel displays the east and
  northeast fronts. Edge locations for the fronts are denoted by a
  vertical dashed line. Uncertainties in the edge fit parameters 
  are colored to correspond to the relevant profiles. Best-fit
  parameters are in Table \protect\ref{tab:cfparam}.}
\label{fig:eastcf}
\end{center}
\end{figure*}

Surface brightness, density, temperature, and pressure profiles for
the west, southeast, east, and northeast elongated edges 
are plotted in Figure \ref{fig:eastcf}. We find the west edge is 
well described by a 
density jump of $1.80^{+0.10}_{-0.08}$ and temperature jump of 
$0.56^{+0.16}_{-0.13}$ located $5\farcs 85^{+0.04}_{-0.03}$~west of
the primary cluster cD galaxy. These measurements yield a pressure jump of 
$1.00^{+0.29}_{-0.24}$ across the edge. We find similar density, 
temperature, and pressure jumps across the southeast, east, and 
northeast edges located at $7\farcs 10^{+0.07}_{-0.03}$, 
$11\farcs 5^{+1.3}_{-1.2}$, and $16\farcs 7^{+0.3}_{-0.5}$~from the 
primary cluster cD galaxy, respectively (see Table \ref{tab:cfparam}). Our density, 
temperature, and pressure ratios show that all of these edges are 
consistent with sloshing cold fronts. Our east and west cold 
fronts agree well with those found by \citet{johnson2012}. Small 
differences in edge radii and jump values are due partly to our 
longer exposure, and also to differences in binning, density model, 
and the geometry and centering used to extract the profiles.

The cold fronts we have measured agree with the edges identified 
visually in the images in \S\ref{sec:asym}. The radial distance 
between the cold fronts and the primary cluster cD galaxy gradually increases 
while rotating clockwise from west. This pattern is characteristic 
of a clockwise gas spiral formed from gas sloshing. Cold fronts 
radially closer to the primary cluster cD galaxy were formed more recently than 
those farther away.  

\begin{figure}[tp!]
\begin{center}
\includegraphics[width=0.495\textwidth]{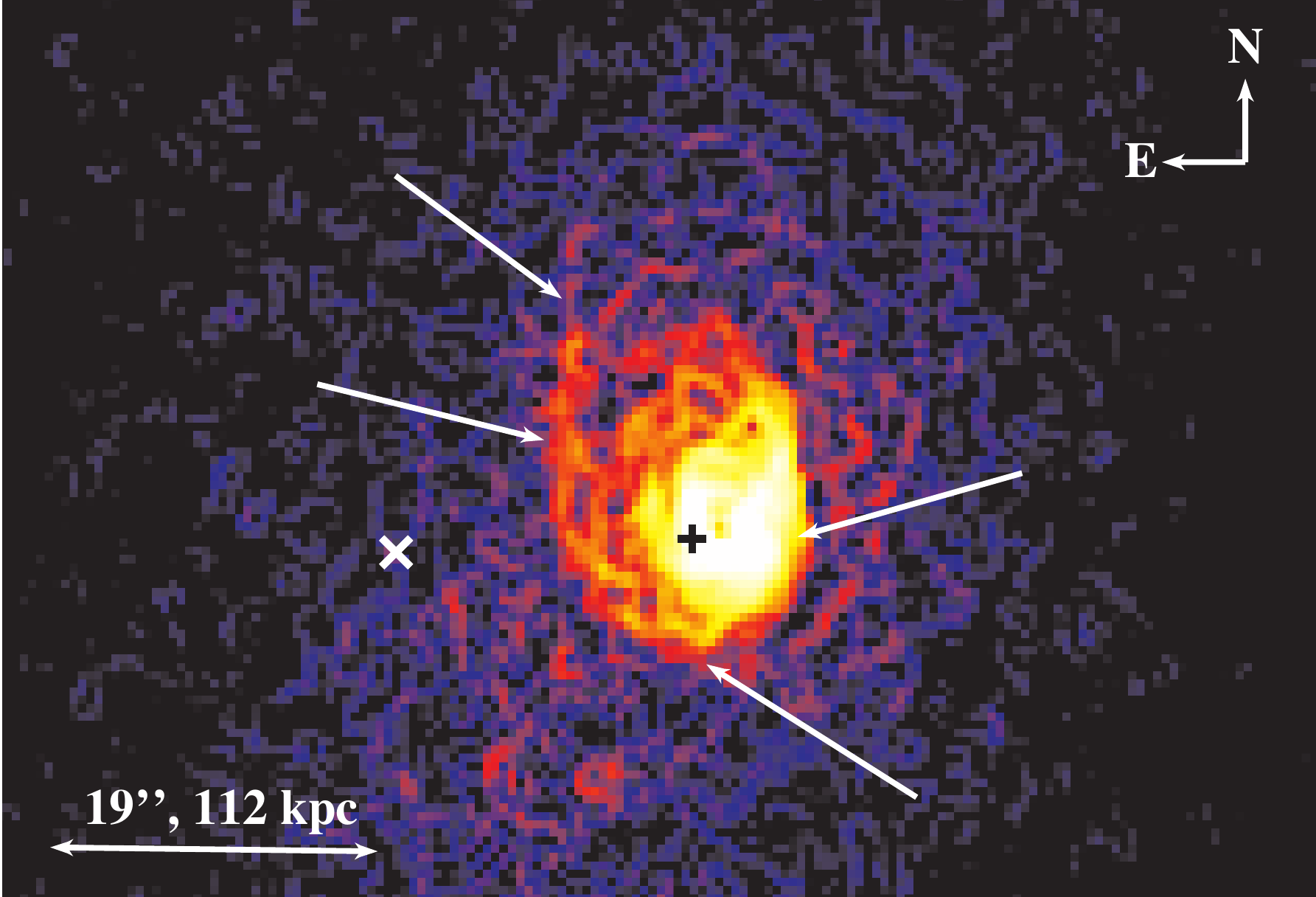} \par
\vspace{0.1cm}
\includegraphics[width=0.495\textwidth]{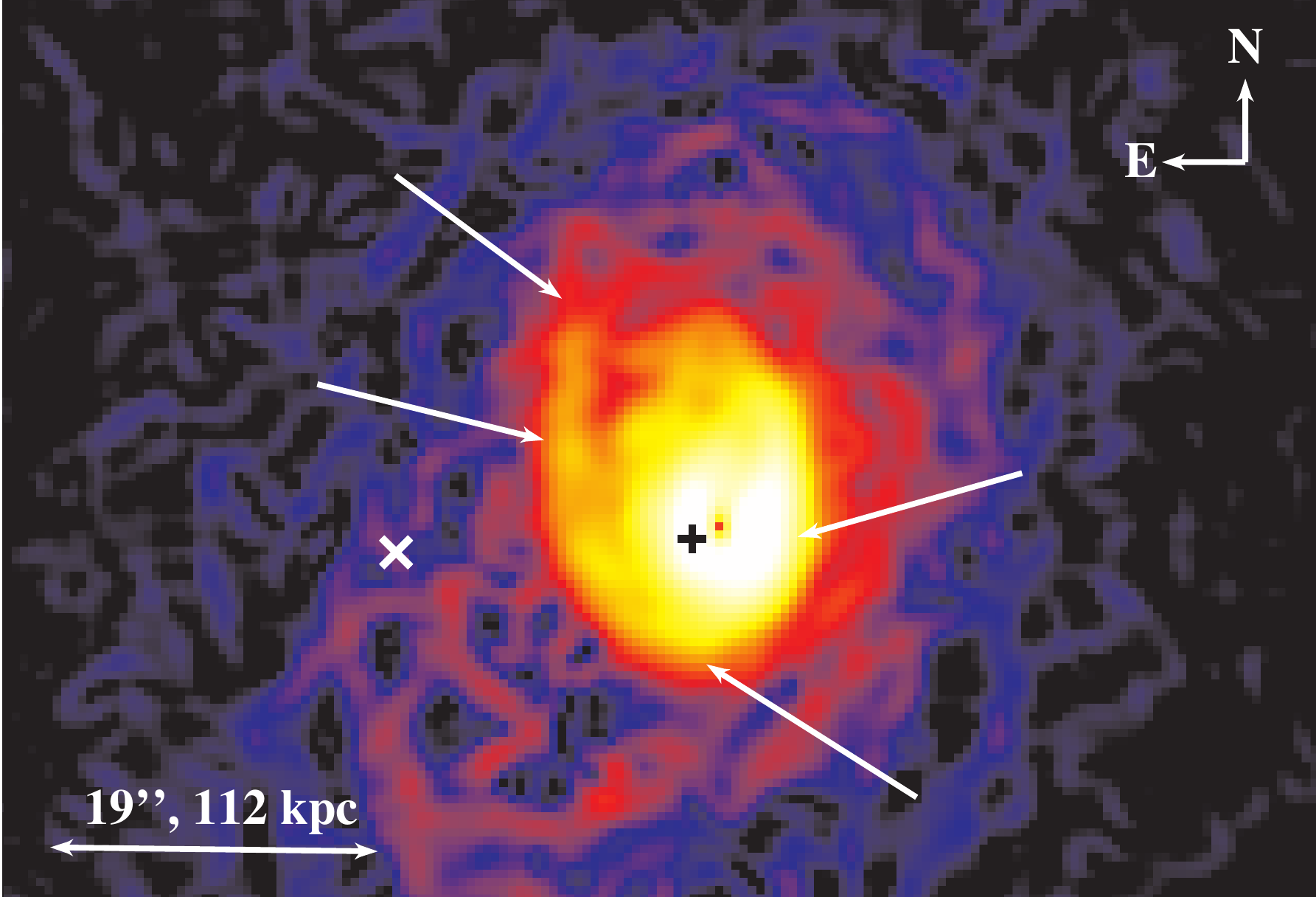} \par
\vspace{0.1cm}
\includegraphics[width=0.495\textwidth]{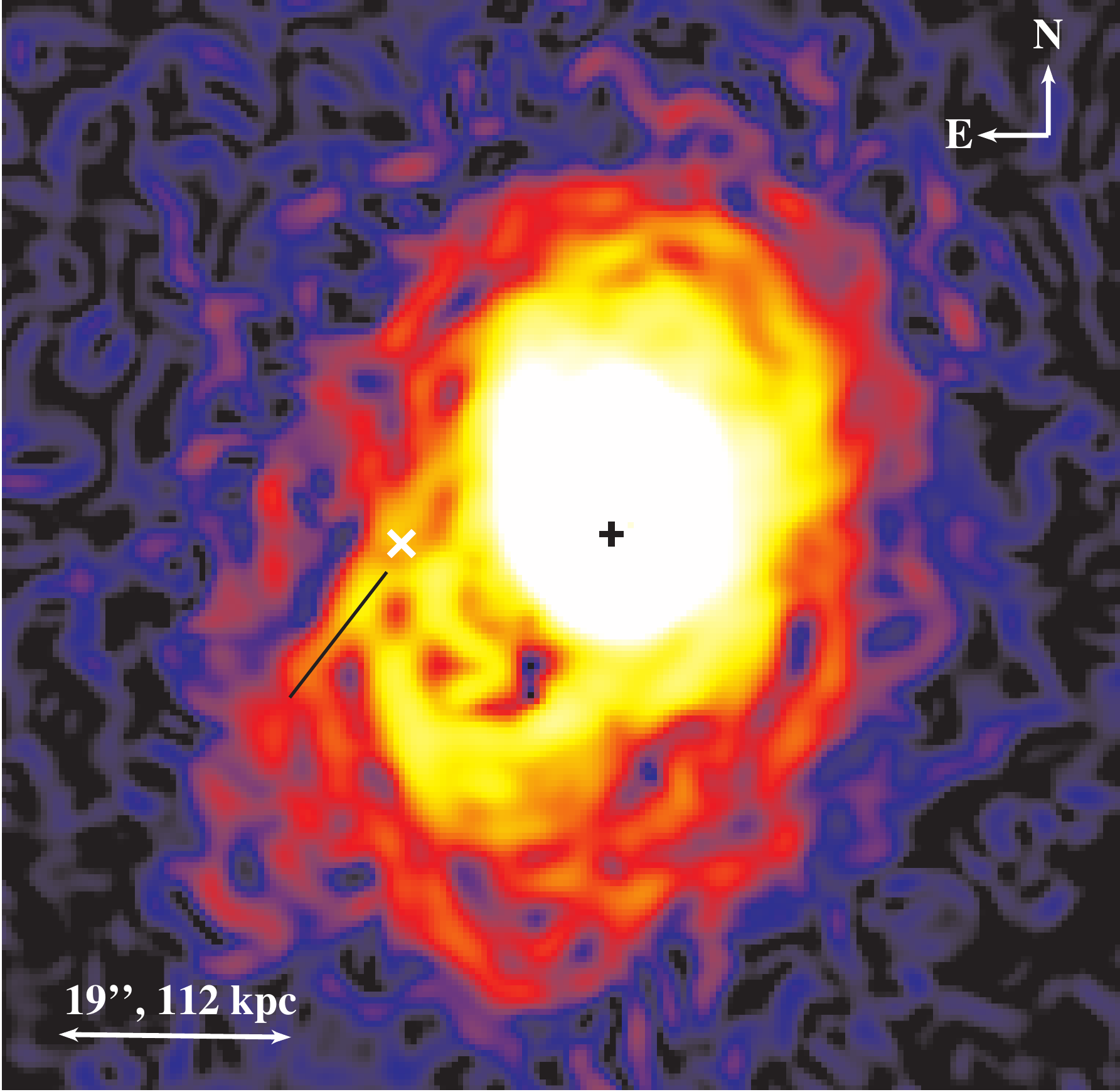}
\caption{GGM filtered images using standard deviations $\sigma$ of 1 (top), 2 (middle), and 3 (bottom). The primary and subcluster BCGs are denoted by a cross and X, respectively. Solid arrows in the top and middle panels point to cold fronts labeled in Figure \ref{fig:xray}. In the bottom panel, the black line corresponds to the ellipse edge that traced the eastern shock front.
}
\label{fig:gradient}
\end{center}
\end{figure}

As an independent check on our edge identifications, we apply the Gaussian gradient magnitude (GGM) filter to the background subtracted, full resolution flux image (Figure \ref{fig:xray}), following the methodology in \citep{sanders2,sanders}. This filter highlights edges by showing high values for large gradients and low values for flat surface brightness profiles. In Figure 
\ref{fig:gradient} we show filtered images using standard deviations $\sigma$ of 1 pixel ($0 \farcs 492$), 2 pixels ($0 \farcs 984$), and 3 pixels ($1 \farcs 476$) in the top, middle, and bottom panels, respectively. A smaller $\sigma$ will highlight small scale gradients, while a larger $\sigma$ will highlight broad features. In the top and middle panels we have overlaid the 4 solid arrows denoting cold fronts seen in Figure \ref{fig:xray}. These arrows coincide with sharp gradients in the map, confirming our edge identifications and clockwise spiral feature. Filtering with $\sigma=1$, the cold fronts appear to be traced by filamentary structures, possibly similar to those in the Centaurus cluster discussed by \citet{sanders}. The bright central region likely corresponds to the bright, cool core in Figures \ref{fig:xray} and \ref{fig:cfpos}. In the middle panel of Figure \ref{fig:gradient} ($\sigma=2$), the cold front spiral is even more prominent. The elbow-like feature seen to the north in the top two panels is likely associated with the cooler, lower entropy northern fan extension noted in Figure \ref{fig:cfpos}.

\section{Subcluster Properties}
\label{sec:subcluster}

\begin{figure*}[tp!]
\begin{center}
\includegraphics[width=0.495\textwidth]{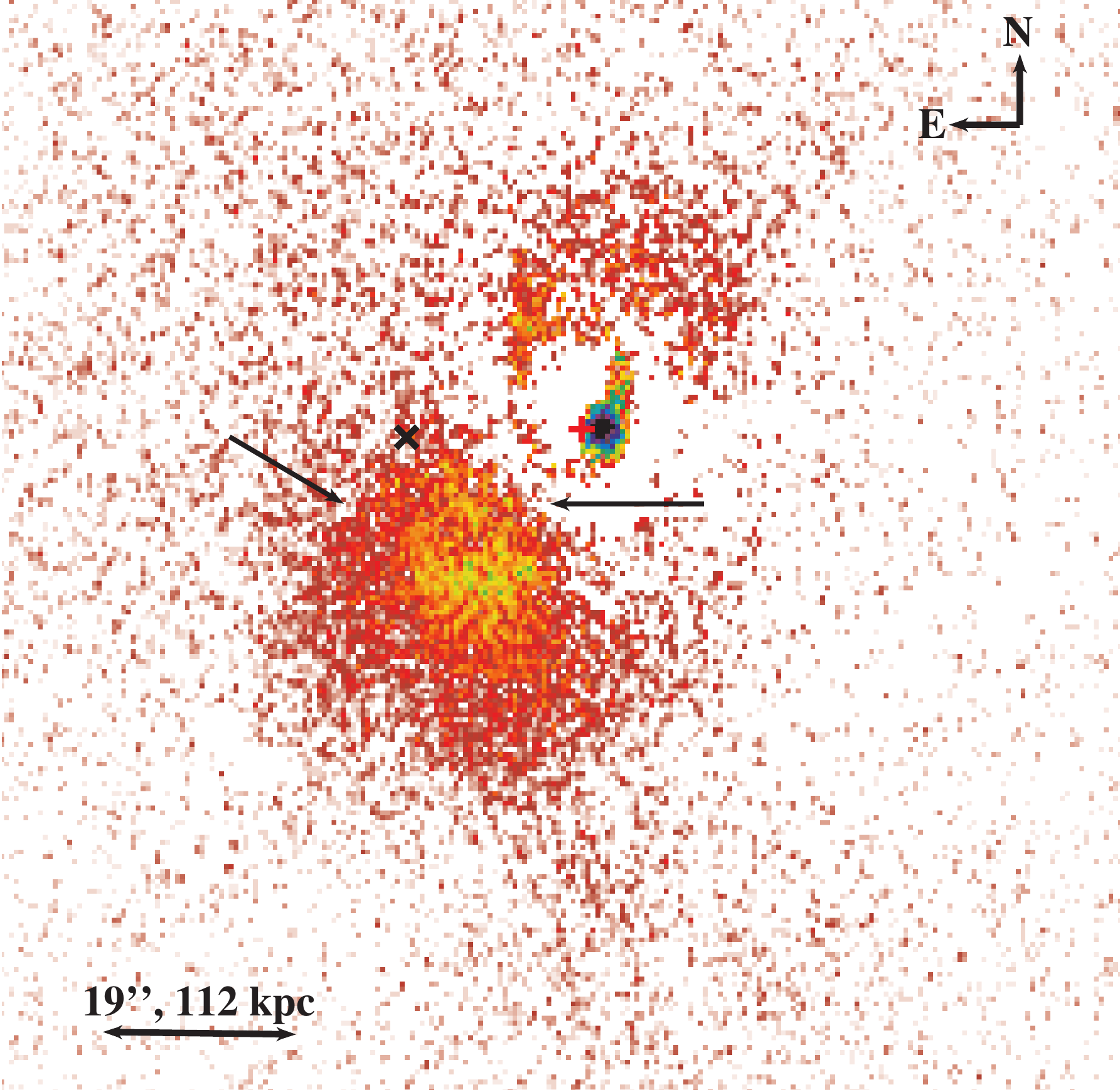}
\includegraphics[width=0.495\textwidth]{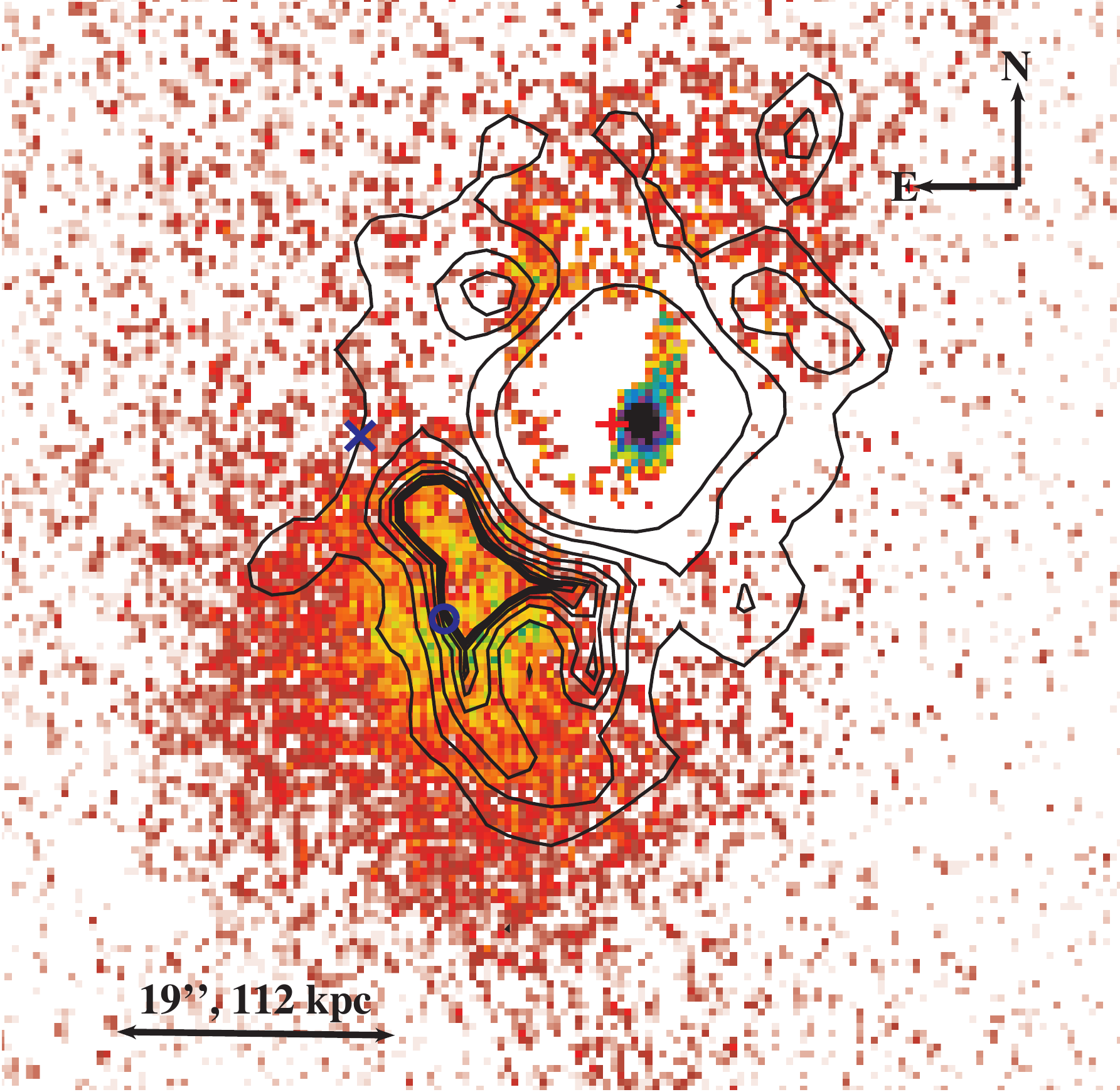}
\caption{(Left) Difference map with log scale resulting from the residuals to
  the spherical beta model fit to the primary cluster surface
  brightness in \S\protect\ref{sec:mod}. Note the prominence of 
  the subcluster. We
  point to the sharp leading edges with arrows. (Right) Difference map 
  with  SZ contours overlaid, indicating the maximum SZ decrement 
  over the western subcluster edge seen in the left panel, suggesting
  the edges represent the Mach cone of a shock front. The primary BCG, subcluster BCG, and subcluster X-ray peak are denoted with a cross, X, and circle, respectively.
}
\label{fig:sub_resid}
\end{center}
\end{figure*}

In Figure \ref{fig:xray} excess X-ray emission, interpreted as gas from a
merging subcluster,  is seen to the southeast of the primary
cluster. 
To better isolate the subcluster and study the  X-ray
morphology of its excess emission, we subtract the mean cluster model
obtained in Section \ref{sec:mod} for
the primary cluster alone from the merged 0.5-2.5 keV X-ray image of the
merging pair. This X-ray surface brightness difference map is given in Figure
\ref{fig:sub_resid}. 

\subsection{Qualitative Features: the Case for Gas Stripping 
and Core Shredding}
\label{sec:sub_qual}

Several qualitative features in Figure \ref{fig:sub_resid} are
striking. The northern edges of the
subcluster appear cone-like with the vertex displaced $\sim 2''$ south of
the second BCG (taken to be the  subcluster's central dominant
galaxy), suggestive of gas displaced from the subcluster gravitational
potential by ram pressure as it infalls through the
primary cluster ICM from the southwest to the northeast. 
The western edge of this cone is in close correspondence to
the maximum SZ decrement, as shown in the right panel of
Figure \ref{fig:sub_resid} \citep{mason2010, korngut2011}, characteristic of
a shock, suggesting that the cone-like feature may be the Mach cone of shocks
formed by the subcluster's supersonic motion through the primary
cluster ICM.  Within the Mach cone is
a bright region suggestive of the subcluster's gas core. The peak of the
X-ray emission, defined as the brightest subcluster pixel in the 
$0.5-2.5$\,keV X-ray image shown in Figure \ref{fig:xray}, is located
at  RA~$13^h 47^m 31.5^s$, Dec $-11^{\circ} 45' 23\farcs 7$. 
The morphology of the excess X-ray emission is elongated
towards the southwest with the hint of an extended tail, also qualitatively
consistent with that expected for subcluster gas being stripped.

\begin{figure*}[tp!]
\begin{center}
\includegraphics[width=0.475\textwidth]{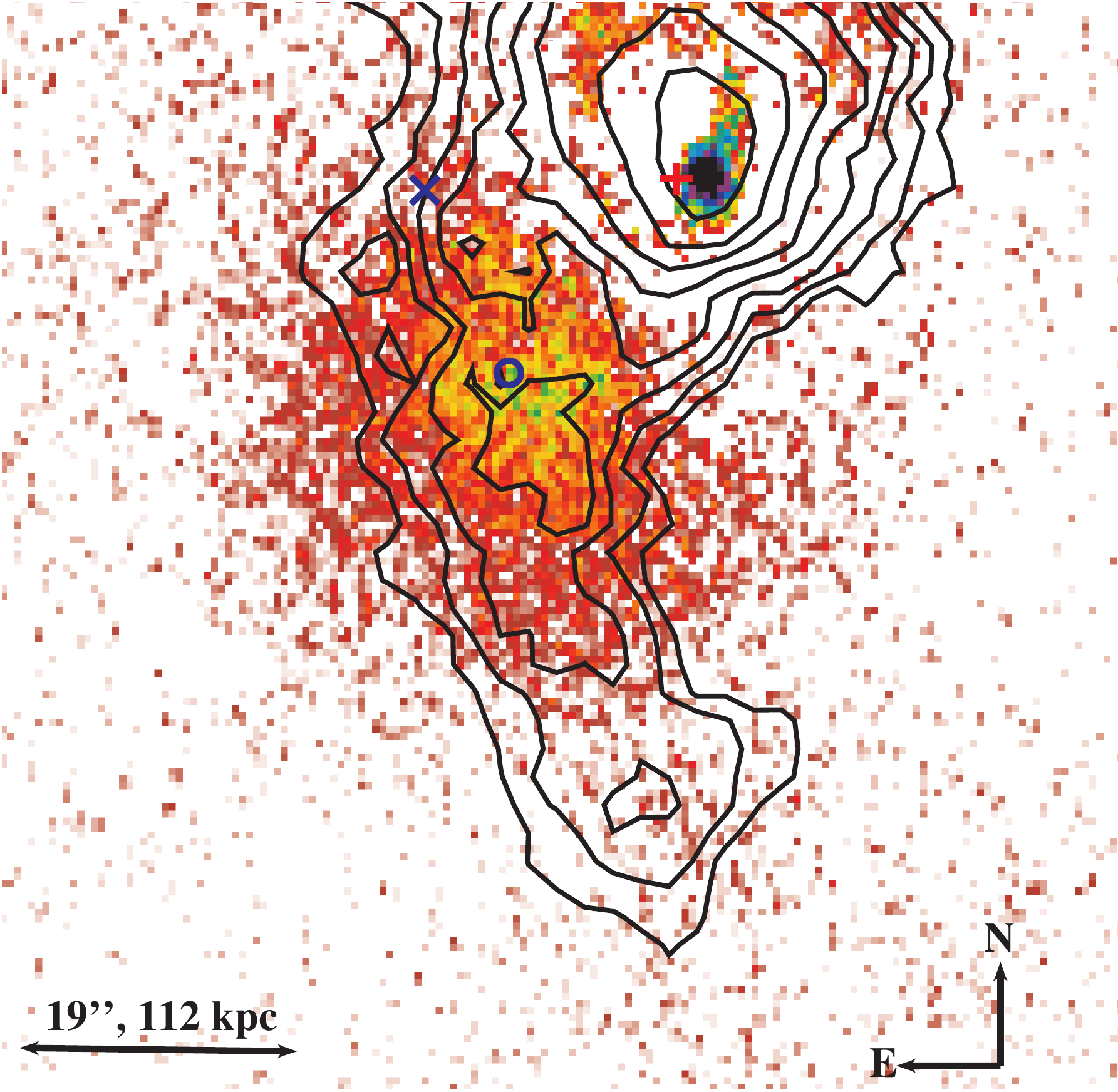}
\includegraphics[width=0.475\textwidth]{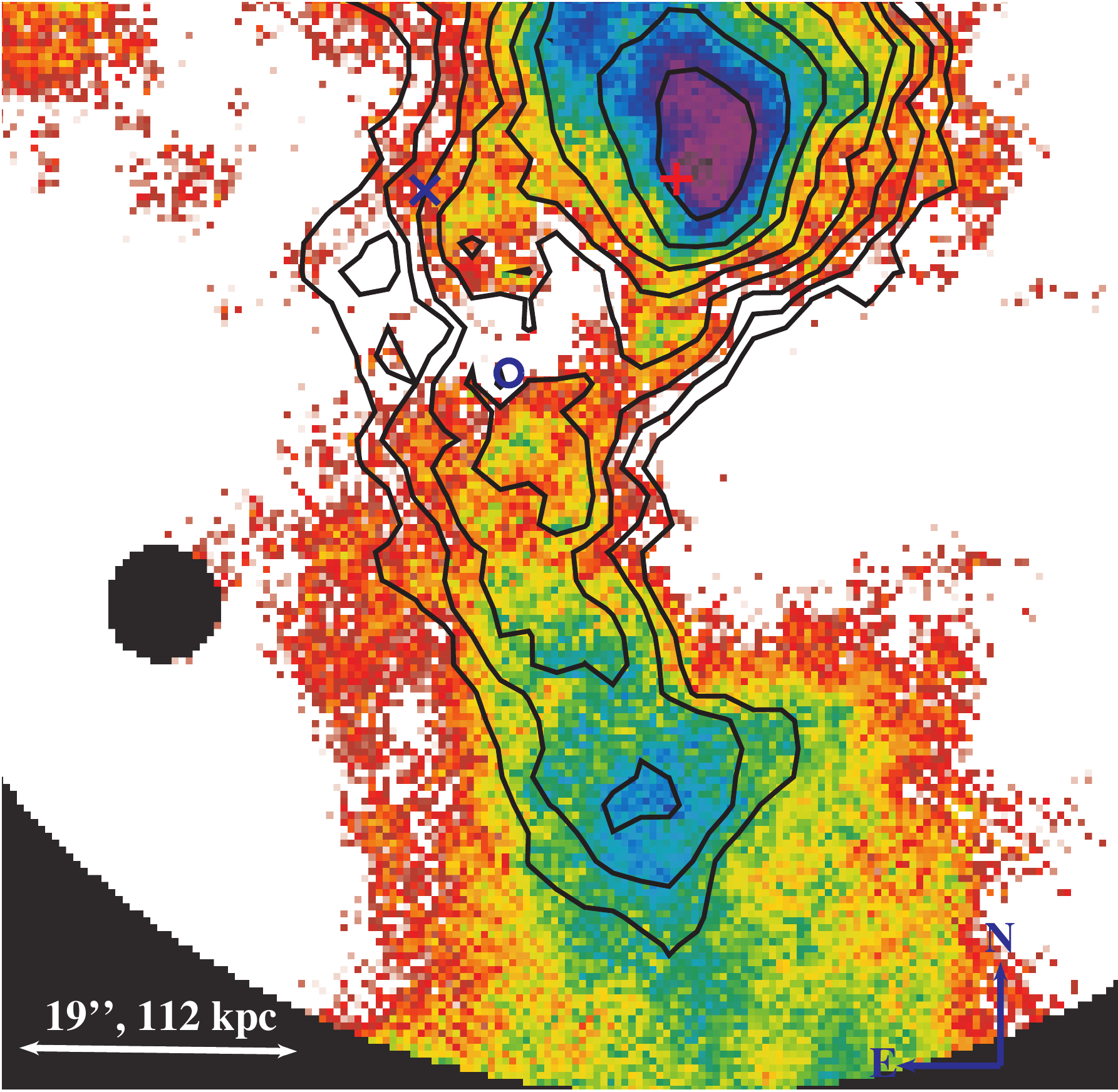}
\caption{The left panel shows our difference map from
  Figure \protect\ref{fig:sub_resid} with entropy contours
  overlaid while the right panel shows the same contours overlaid on
  the temperature map
  for the same region matched in WCS coordinates. The primary BCG,
  subcluster BCG, and subcluster X-ray peak are denoted with a red
  cross, blue X, and blue circle, respectively. The black circle in
  the temperature map is a removed point source. 
}
\label{fig:entropy}
\end{center}
\end{figure*}

\begin{figure*}[bp!]
\begin{center}
\includegraphics[width=0.475\textwidth]{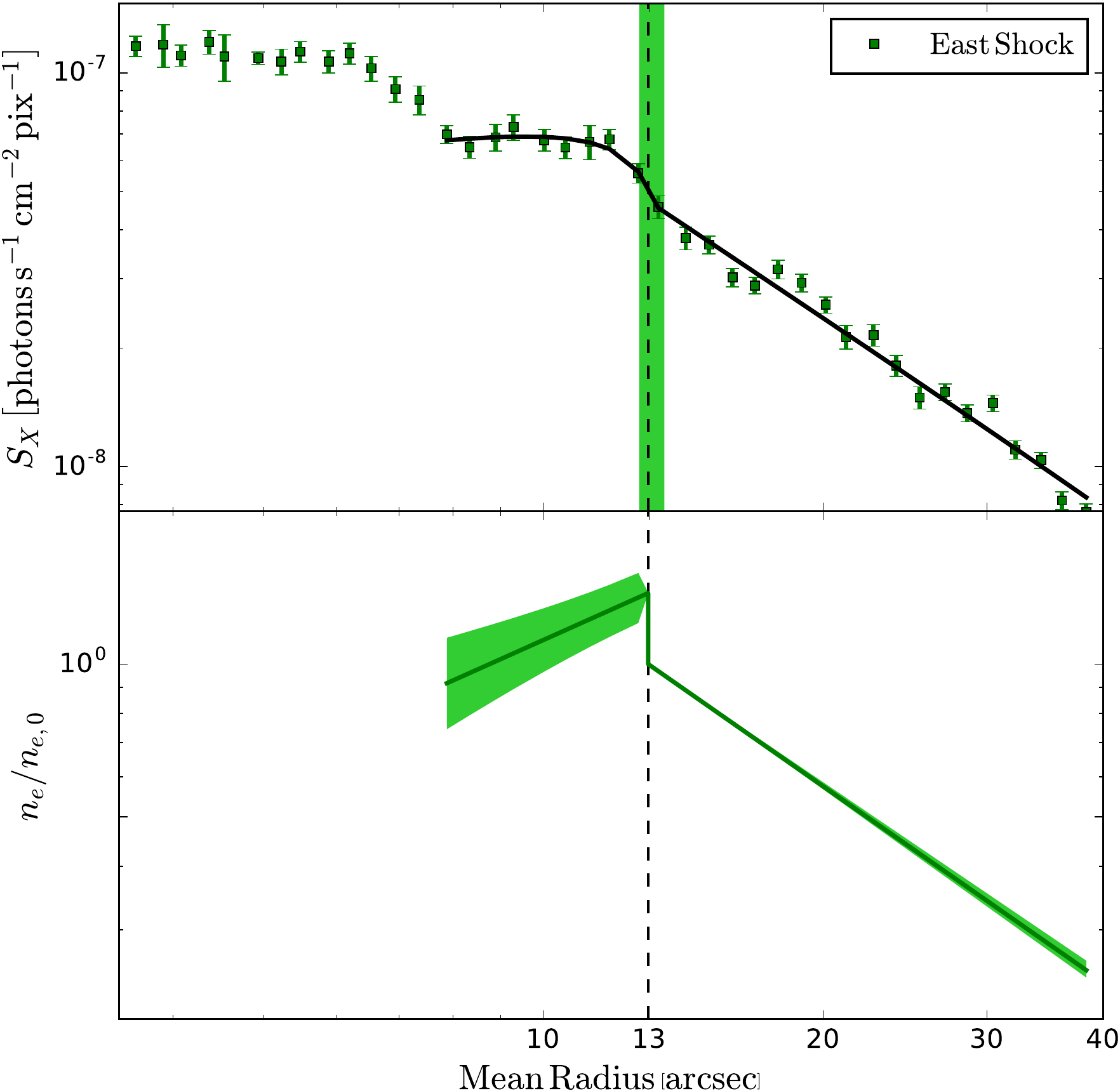}
\includegraphics[width=0.475\textwidth]{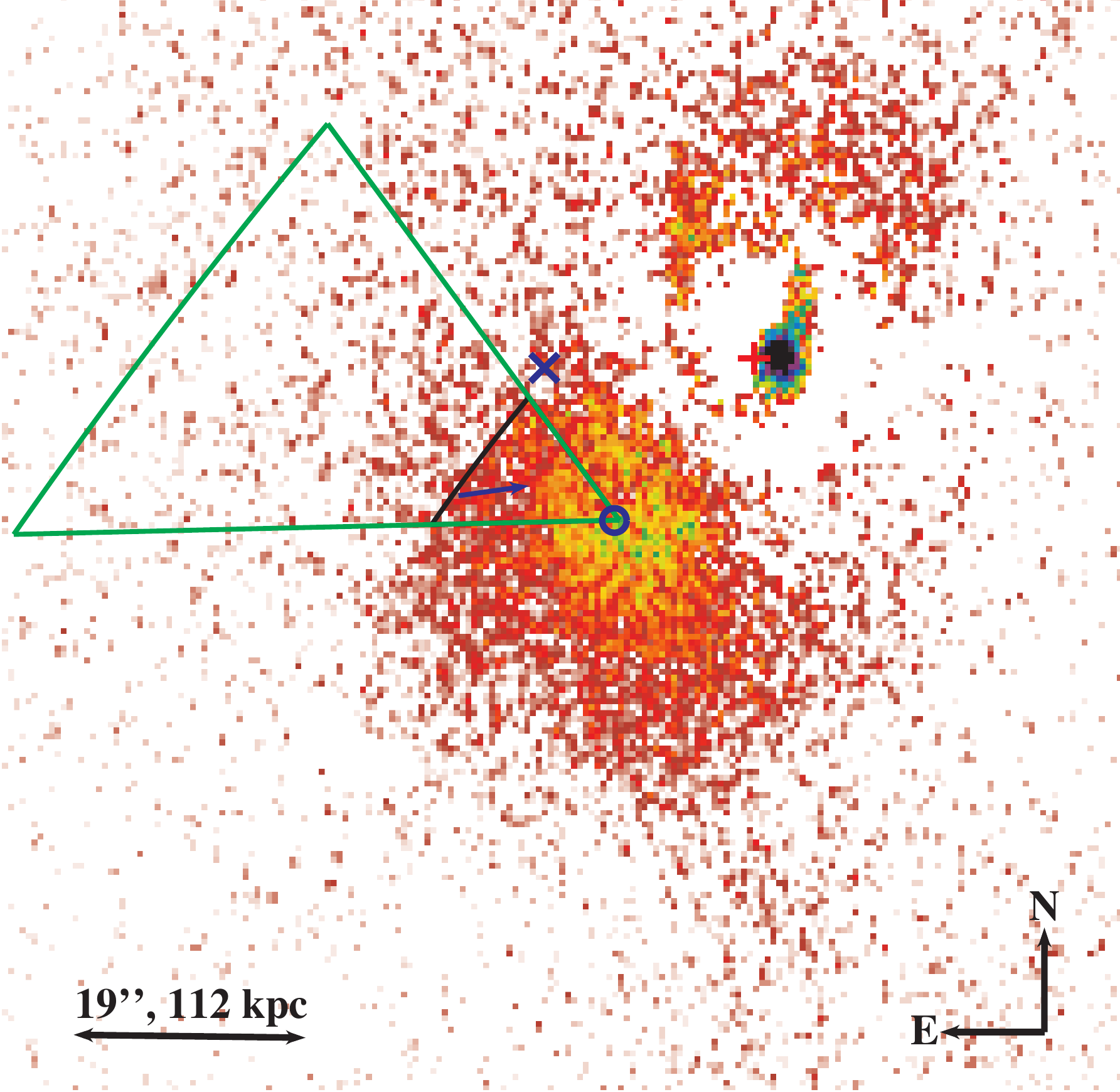}
\caption{(Upper Left) The surface brightness profile across the
  eastern shock front. The black line shows the best fit assuming a 
broken power law (Equations \ref{eq:cf1} and \ref{eq:cf2}) for the gas
density. The potential merger cold front surface brightness jump is 
seen at~$\sim7-8''$. (Lower Left) Best fit gas density from the upper 
panel with power law
indices~$-1.28^{+0.04}_{-0.03}$~($0.82^{+0.70}_{-0.59}$) 
upstream (downstream) of the shock. The shock edge location is denoted 
with a dashed line and $90\%$~CL uncertainties shaded in
green. (Right) The sector used to construct the surface brightness
profile across the shock is overlaid on the residuals image with log
scale. The black line shows the segment of the bounding ellipse that 
traces the eastern shock edge within the sector. The potential merger 
cold front surface brightness discontinuity is denoted with a blue arrow. See also the bottom panel of Figure \ref{fig:gradient}.}
\label{fig:sfell}
\end{center}
\end{figure*}

In Figure \ref{fig:entropy} we show pseudo-entropy contours overlaid on the
surface brightness difference map (left panel) and temperature map
(right panel) for the subcluster gas. Low entropy contours coincide
with luminous, lower temperature regions in the subcluster and
tail. We locate a `ledge' in the southern contours,
indicating an irregular distribution of low entropy gas in the
elongated region. In a cluster whose core is still intact, we would
expect a spherical entropy distribution with regular, monotonically
increasing contours of entropy as a function of distance from the
subcluster gas peak. While there is a concentration of low entropy gas
near the subcluster X-ray surface brightness peak, the elongated,
asymmetric entropy contours and lower temperatures to the southwest
of the subcluster peak suggest ongoing core shredding as cooler core
gas gas is pushed back towards the tail-like structure. These qualitative features are confirmed by the GGM $\sigma=3$ map shown in the bottom panel of Figure \ref{fig:gradient}. The eastern shock edge in the GGM image corresponds to the edge ellipse used for our shock density ratio analysis in \S \ref{sec:shocks} (see Figure \ref{fig:sfell}). The western gradient edge corresponds to the SZ decrement. Together they trace the shock Mach cone. The bright extension to the southwest likely corresponds to the subcluster's stripped gas tail.

\begin{deluxetable}{ccc}
 \tablewidth{0pc}
 \tablecaption{Circular Subcluster and Background Regions\label{tab:subreg}}
 \tablehead{\colhead{Region} & \colhead{Center} & \colhead{Flux ($0.5-2.5$\,keV)
     } \\
      & $\left( \mbox{RA}\right)$, $\left( \mbox{Dec}\right)$ & $10^{-4}$\,\photflux}
 \startdata

  SC & $13:47:31.6$, $-11:45:27.4$ & $1.716\pm 0.016$ \\
 BE  & $13:47:32.1$, $-11:45:02.2$ & $0.775 \pm 0.011$ \\
 BN  & $13:47:30.4$, $-11:44:48.7$ & $1.491 \pm 0.015$ \\
 BW  & $13:47:29.2$, $-11:45:06.9$ & $0.634 \pm 0.010$ 
 \enddata
 \tablecomments{SC includes emission from both the primary cluster and
   subcluster at $22''$ from the primary cluster central
   galaxy. BE, BN, and BW denote circular regions at the
   same radial distance containing primary cluster emission only. All
   circular regions have radius $11\farcs 7$. Flux uncertainties are $1\sigma$.
 }
 \end{deluxetable}

\subsection{Subcluster Temperature, Luminosity, and Mass}
\label{sec:mass}

\begin{figure}[tp!]
\begin{center}
\includegraphics[width=0.5\textwidth]{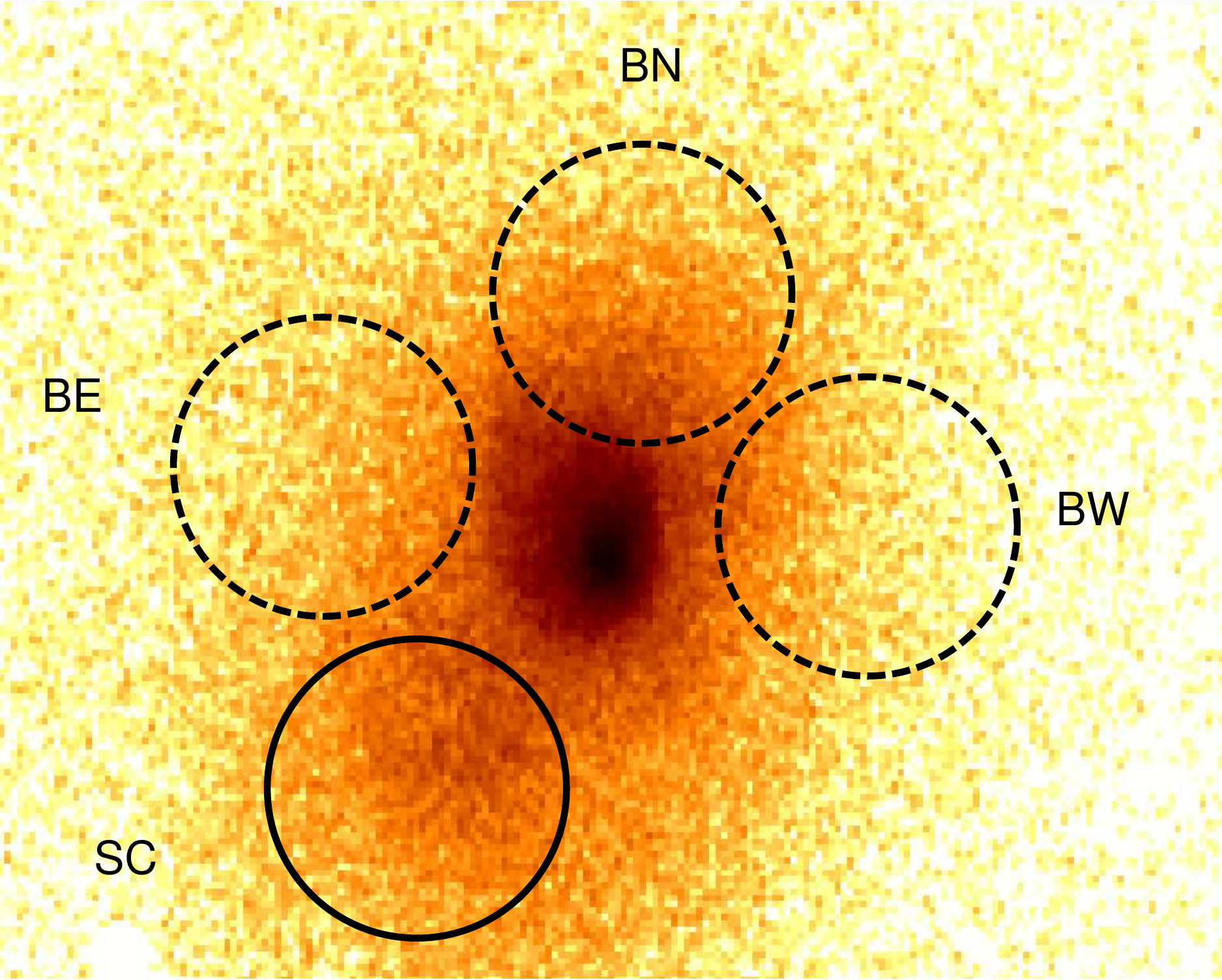}
\caption{Source (subcluster + cluster, solid circle)  and 
local background (cluster only, dashed circles) regions overlaid on
the $0.5-2.5$\,keV flux image of RXJ1347. Note the excess emission in
background region BN due to the northern extension of the spiral and
fan. 
}
\label{fig:sub_spec_reg}
\end{center}
\end{figure}

To measure the temperature and luminosity of the subcluster, 
we extract the subcluster spectrum from a circular region of radius
$11 \farcs 7$ located $22''$ from the primary cluster central galaxy, that
covers the majority of the subcluster emission. 
To minimize the contribution to the spectrum from the primary cluster gas, we choose local background circles of
the same size and distance from the primary cluster central
galaxy as the subcluster, as in \citet{johnson2012}. However, the sloshing features (spiral and fan) cause the flux
to vary with angle by as much as a factor of
$2$ at this distance (see Table \ref{tab:subreg} and Figure \ref{fig:sub_spec_reg}). We argue that the X-ray flux in the
northern local background circle (BN in Figure \ref{fig:sub_spec_reg}) is 
anomalously high compared to that expected at the location of the 
subcluster due to this excess gas displaced to the north by sloshing. These sloshing features found to
the north would not be expected at that radius in the southeast at the
location of the subcluster. We instead fit the subcluster
spectrum using the combined eastern (BE) and western (BW) cluster
background regions as more representative of cluster emission at 
the location of the subcluster. 

\begin{deluxetable}{ccccccc}
 \tablewidth{0pc}
 \tablecaption{Subcluster Temperatures\label{tab:subtemp}}
 \tablehead{\colhead{Region} & \colhead{Temperature} & \colhead{$\chi^2$/(d.o.f.)}\\
       & $\left( \mbox{keV}\right)$ & }
 \startdata
  Total Subcluster Gas & $16^{+4}_{-3}$ & $621.66/588$ \\
  SZ Decrement & $19^{+6}_{-4}$ & $164.74/153$ 
 \enddata
 \tablecomments{Temperatures are presented with $90 \%$~CL after being 
fit with an absorbed APEC model with fixed Galactic absorption and abundance
at~$0.3\,\Zs$. 
 }
 \end{deluxetable}

Using an absorbed APEC thermal plasma model and fixing the absorbing column at the Galactic value ($4.75\times 10^{-20}\,{\rm cm}^{-2}$), we find a subcluster temperature of $kT=16^{+4}_{-3}$\,keV (see Table \ref{tab:subtemp}) and abundance $A=0.5^{+0.3}_{-0.2}\,\Zs$ ($\chi^2/{\rm dof}=629/587$). From directly summing the $0.5-2.5$\,keV flux in the source and subtracting the expected primary cluster contribution using the BE+BW background regions, we measure an absorbed $0.5-2.5$\,keV flux for the subcluster alone of $1.01 \pm 0.02 \times 10^{-4}$\photflux ($1.96 \pm 0.03 \times 10^{-13}$\ergscm ). While these $1\sigma$ statistical uncertainties are small given the long exposure, we caution the reader that the uncertainties in measuring the subcluster flux are dominated by the choice of model for the primary cluster emission at that location, and may be as high as $\sim 25\%$.
 
Our mean value for the net absorbed $0.5-2.5$~keV flux in the
subcluster is higher (but still within their $1\sigma$ uncertainties) 
of that previously obtained by \citet{johnson2012} 
($1.5 \pm 0.9 \times 10^{-13}$\ergscm), because we
exclude the northern region (BN) from our estimate of the  average 
primary cluster emission and they did not. 
Using the spectral model to account for Galactic
absorption and modest change in energy band to be consistent with 
\citet{zhang2011},
we find the unabsorbed $0.5-2$\,keV X-ray 
luminosity of the subcluster to be  
$L_{\rm X}= 1.3 \times 10^{44}$\ergs.
Applying the
$M_{gas,500}-L_{0.5-2keV,500}$ relationship from \citet{zhang2011} 
for disturbed clusters including the core correction,   

\begin{eqnarray}
\log_{10}(Y)=A+B\log_{10}(X)
\end{eqnarray}
where
\begin{eqnarray}
Y=\frac{L_{0.5-2keV,500}}{E(z)~\mbox{erg~s$^{-1}$}}
\end{eqnarray}
\begin{eqnarray}
X=\frac{M_{gas,500}E(z)}{10^{14}~\mbox{M$_{\odot}$}}
\end{eqnarray}
with  $A=44.44 \pm 0.61$, $B=1.16\pm 0.04  $~at~$1\sigma$, and 
\begin{eqnarray}
E^2(z)=\Omega_m(1+z)^3+\Omega_{\Lambda}+(1-\Omega_m-\Omega_{\Lambda})(1+z)^2
\end{eqnarray}
we calculate the enclosed gas mass to be~$3.3 \times10^{13}\,\Ms$.  
Assuming the gas fraction is $10 \%$ of the total 
mass \citep{vikhlinin2006}, this yields the subcluster's total mass to 
be~$3.3 \times10^{14}$~M$_\odot$, consistent with the mass estimate found 
by \citet{johnson2012}. Since we expect that some subcluster gas may
already have been stripped due to the encounter, this should be interpreted
as a lower bound on the subcluster total mass.   
  
\begin{deluxetable*}{ccccccccc}
 \tablewidth{0pc}
 \tablecaption{Eastern Shock Properties \label{tab:sfparam}}
 \tablehead{\colhead{Location of Edge} &
   \colhead{$\frac{\rho_{2,m}}{\rho_{1,m}}$} & \colhead{$\frac{T_2}{T_1}$} & 
   \colhead{$\frac{P_2}{P_1}$} & \colhead{Mach number} & \colhead{Velocity} \\
     $\left( \mbox{arcsec}\right)$ & & & & & $\left( \mbox{km\,s}^{-1}\right)$ }
 \startdata
  $13.0^{+0.5}_{-0.3}$ & $1.38^{+0.16}_{-0.15}$ & $1.25$ & $1.72$ 
   & $1.25 \pm 0.08$ & $2810^{+210}_{-240}$
 \enddata
 \tablecomments{Shock properties were obtained by fitting a radial
   profile centered on the subcluster X-ray peak. Subscripts of $2$
   and $1$ denote quantities downstream and upstream of the shock, 
respectively, whereas the subscript~$m$ denotes a measured, 
rather than derived, quantity. $T_2/T_1$, $P_2/P_1$, and the Mach
number were derived from the Rankine-Hugoniot jump conditions using
the measured density ratio. The shock velocity assumes an 
unshocked gas temperature of $19\pm3$\,keV.}
 \end{deluxetable*}

\subsection{Eastern Shock Front and Shock Velocity}
\label{sec:shocks}

Following the same procedure as in \S\ref{sec:coldfront} with 
the~$0.5-2.5$~keV full resolution exposure-corrected X-ray flux image, 
we construct the surface brightness profile across the eastern shock 
edge using elliptical wedge regions concentric to a bounding ellipse 
tracing the eastern shock edge and constrained to lie within the 
sector subtending the angles from~$126^{\circ}$~to~$181^{\circ}$. The 
bounding ellipse is centered on the subcluster X-ray peak with 
semi-major (-minor) axes and position angle of~$44''$ ($12''$) 
and~$51^{\circ}$, respectively (see Figure \ref{fig:sfell}). We model
the gas density using a broken power law density model across the
subcluster eastern edge (see Equations \ref{eq:cf1} and \ref{eq:cf2}, 
\S\ref{sec:coldfront}). To fit the surface brightness profile shown 
in the upper left panel of Figure \ref{fig:sfell}, we integrate the
square of the density model along the line of sight by using a 
multivariate~$\chi^2$~minimization scheme. Our results for the best
fit subcluster gas density ratio across the shock are shown in the 
lower left panel of Figure \ref{fig:sfell}. We find the shock edge is 
located~$13\farcs 0^{+0.5}_{-0.3}$~northeast of the subcluster X-ray
peak with a density ratio of~$1.38^{+0.16}_{-0.15}$. 

The shock Mach number can be determined from either the density ratio 
$\rho_2/\rho_1$ or temperature ratio $T_2/T_1$ across the shock using  
the Rankine-Hugoniot jump conditions 
\citep{RK,korngut2011}
\begin{eqnarray}
\label{eq:rk1}
\mathcal M_{\rho} = \left[\frac{2\frac{\rho_2}{\rho_1}}{\gamma+1-\left(\gamma-1\right)\frac{\rho_2}{\rho_1}}\right]^{1/2}
\end{eqnarray}
and 
\begin{eqnarray}
\label{eq:rk2}
\mathcal M_T =
\left\{\frac{8\frac{T_2}{T_1}-7+\left[\left(8\frac{T_2}{T_1}-7\right)^2+15\right]^{1/2}}{5}\right\}^{1/2}
\end{eqnarray}
where $\gamma=\frac{5}{3}$, and $\rho_2$ ($T_2$) and $\rho_1$ ($T_1$)
are the gas densities (temperatures) inside and outside the shock,
respectively. Using our measured density ratio in
Equation \ref{eq:rk1}, we find a shock Mach number 
$M_{\rho} =1.25 \pm 0.08$. 

We calculate the velocity of the eastern shock by multiplying the Mach
number by $c_s$, the speed of sound in the unshocked primary cluster 
plasma, using
\begin{eqnarray}
c_s^2=\frac{\gamma P_g}{\rho_g}
\end{eqnarray}
where
\begin{eqnarray}
P_g=n_gk_BT_K
\end{eqnarray}
\begin{eqnarray}
\rho_g=n_g\mu m_p
\end{eqnarray}
and $\gamma=\frac{5}{3}$ and $\mu=0.6$. Thus,
\begin{eqnarray}
c_s^2=\frac{\gamma k_BT_K}{\mu m_p} & = & \frac{\gamma T_{eV}}{\mu m_p}
\end{eqnarray}
and
\begin{eqnarray}
\label{eq:soundspeed}
c_s=516~\mbox{km \, s$^{-1}$} T_{keV}^{1/2}.
\end{eqnarray}

Using the local background regions from Table \ref{tab:lbgreg}, we extract 
a spectrum for region BE from Figure \ref{fig:sub_spec_reg} and 
Table \ref{tab:subreg}
to the north of the tip of the Mach cone (see Figure \ref{fig:sub_resid}) 
to measure the temperature and 
determine the local sound speed in the cluster gas outside the shock. 
Modeling the spectrum with an absorbed APEC model with fixed Galactic
absorption and abundance at $0.3\,\Zs$, we find a cluster gas
temperature $kT=19\pm3$\,keV ($\chi^2/{\rm dof}=297/305$ for $7386$ net
source counts, see Table \ref{tab:subtemp}). 
From Equation \ref{eq:soundspeed}, we 
estimate the speed of sound in the unshocked cluster gas 
as $c_s=2250^{+170}_{-190}$\kms and eastern shock velocity of
$2810^{+210}_{-240}$\kms.

From Equation \ref{eq:rk2},  we predict a
(deprojected) temperature ratio $T_2/T_1 = 1.25$.    
To test this prediction, one would ideally extract spectra in narrow
regions on either side of the shock edge. However, we cannot measure 
the temperature jump in such regions to within $25 \%$ uncertainty with 
these data ($T_2=16^{+23}_{-7}$ keV and $T_1=14^{+26}_{-6}$ keV, yielding $T_2/T_1 = 1.2^{+2.7}_{-0.7}$). Although we find no significant difference between the temperatures
of the subcluster and the cluster gas just outside the shock, the measured temperature ratio is consistent with the Rankine Hugoniot 
prediction within $90\%$ CL uncertainties. 

\subsection{Subcluster Cold Front?}
\label{sec:subcoldfront}

In the elliptical surface brightness profile across the eastern 
shock (Figure \ref{fig:sfell}), we find a second jump in the surface 
brightness $\sim 5''$ closer to the subcluster X-ray peak than the 
shock front. The surface brightness discontinuity is small and gas
 temperatures are high and have a complex distribution, so
 measurements are highly uncertain, causing us to not complete a 
formal edge analysis. However, qualitatively the double-jump nature 
of this profile is reminiscent of the Bullet subcluster 
\citep[see][]{bullet}. For the Bullet, the inner jump was a cold front 
between the subcluster core and outer gas. The concentration of low 
entropy gas near the X-ray peak for our subcluster suggests that 
there is still a remnant core, and that, as in the Bullet cluster, 
the inner surface brightness edge may correspond to a merger cold front. 

\begin{figure*}[tp!]
\begin{center}
\includegraphics[width=0.475\textwidth]{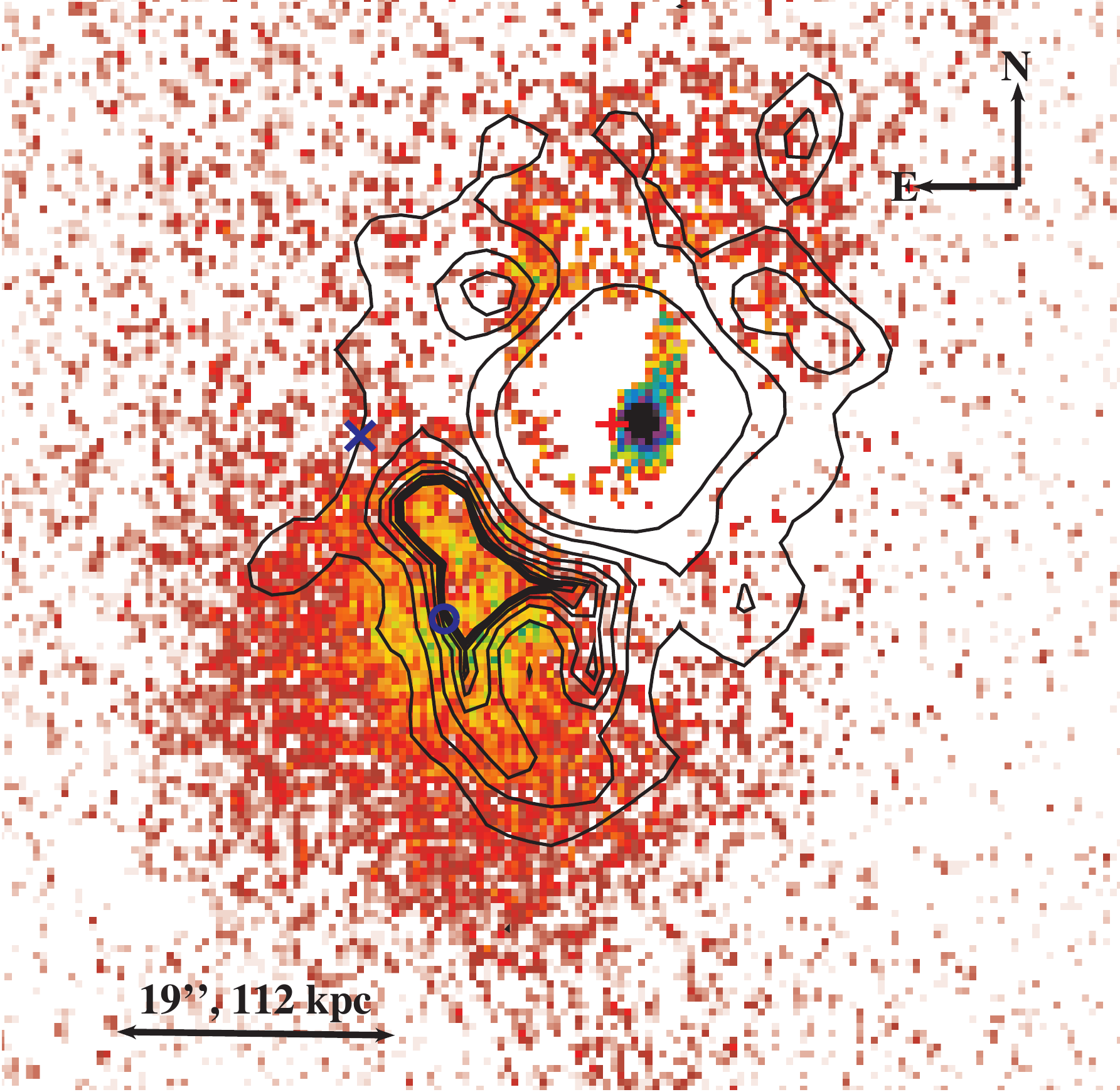}
\includegraphics[width=0.475\textwidth]{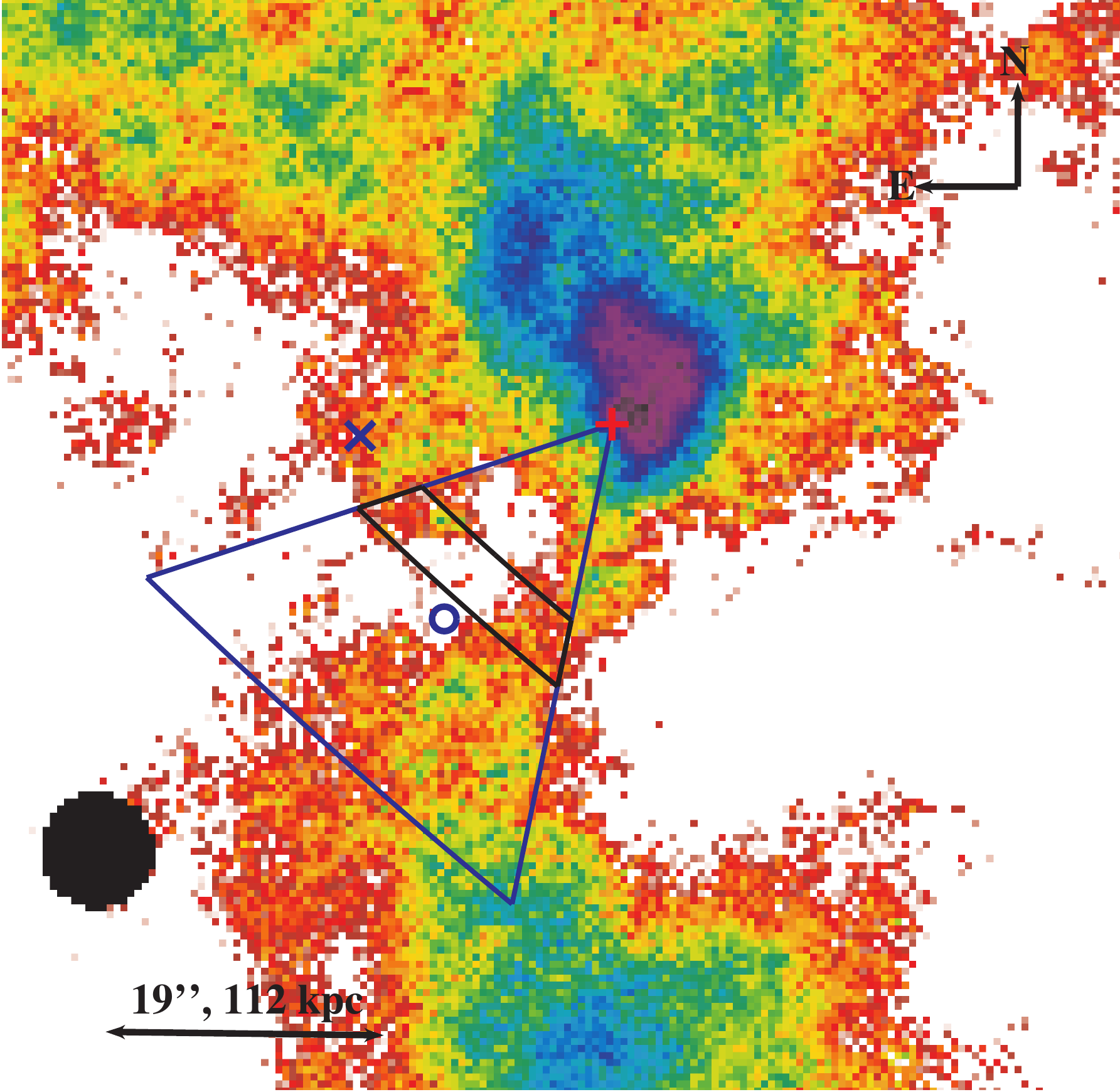}
\caption{(Left) Residuals in log scale with SZ contours overlaid, 
indicating the maximum SZ decrement over the western shock. 
(Right) Projected temperature map in linear scale with the elliptical
 profile covering the western shock front. High gas temperatures 
coincide with the western and eastern shocks, while lower gas
temperature are found in the subcluster core and tail. The region 
enclosed in black corresponds to the gas build up at the shock front 
seen in the surface brightness profile, shown in 
Figure \protect\ref{fig:west_sfell}.}
\label{fig:resid}
\end{center}
\end{figure*}

\begin{figure}[bp!]
\begin{center}
\includegraphics[width=0.5\textwidth]{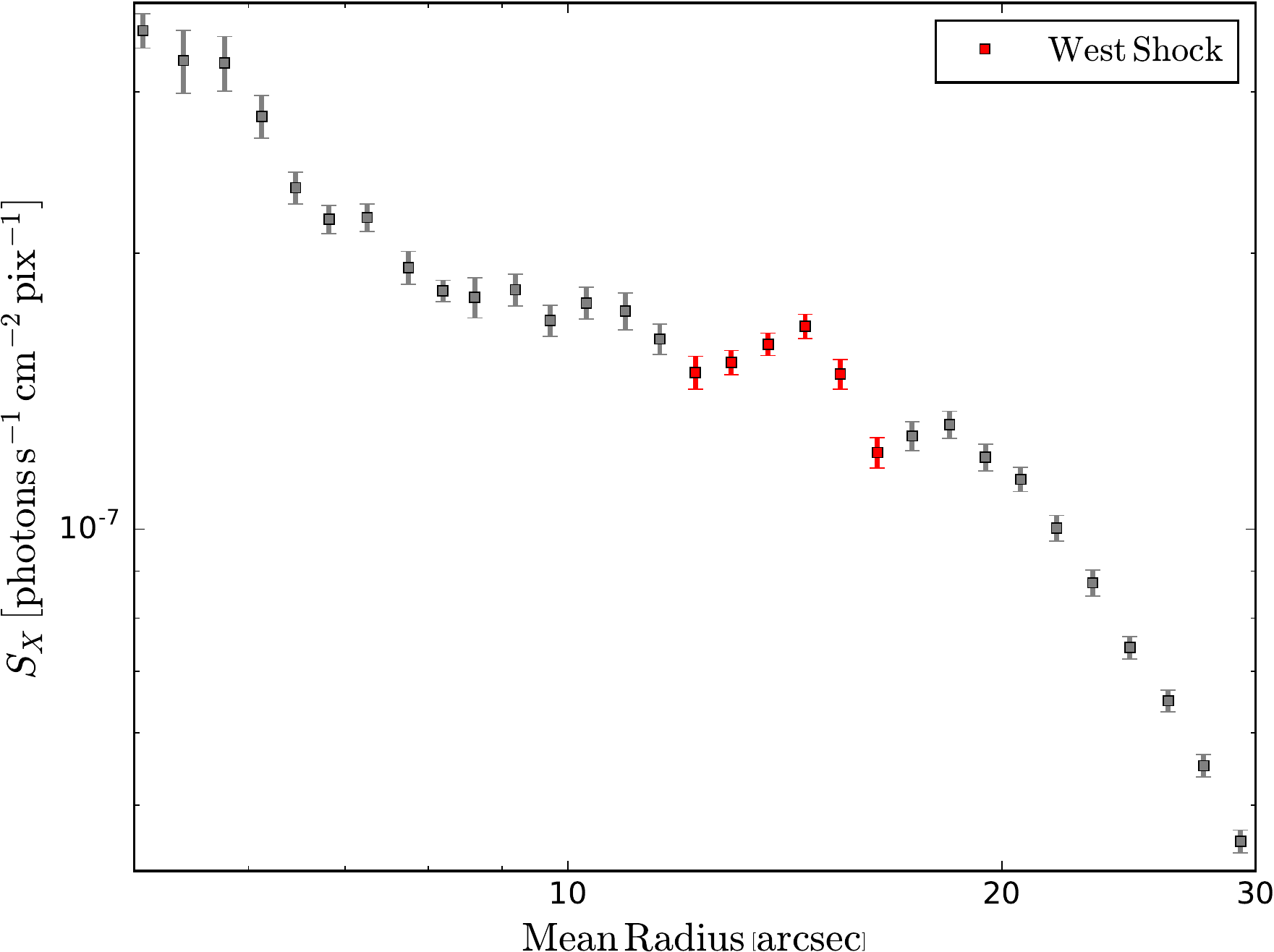}
\caption{Elliptical surface brightness profile centered on the primary 
cD and aligned with the western shock front. The pile up of gas over 
the front is highlighted in red. The red points correspond to the 
region enclosed in black in Figure \protect\ref{fig:resid}.
}
\label{fig:west_sfell}
\end{center}
\end{figure}

\subsection{Western Shock Front}

As the subcluster merges with the primary cluster, gas is shock
heated. High temperatures southwest of the subcluster cD galaxy suggest 
the presence of a western shock front (see right panel of 
Figure \ref{fig:resid}). SZ contours from a MUSTANG SZ map, provided 
by \citet{mason2010} and smoothed by a $10''$ FWHM Gaussian, were 
overlaid on the difference map (left panel of Figure \ref{fig:resid}). 
The most significant SZ decrement over the subcluster occurs from
$\sim13''$ to $\sim17''$ southeast of the primary cluster cD galaxy. We choose a
rectangular region centered at RA, Dec\,$=13^h47^m31.4^s, -11^\circ45'19\farcs5$
with long (short) sides of $12\farcs 3$ ($4\farcs 5)$, respectively, oriented at
position angle $130^\circ$ along the most significant SZ decrement to study
this region. Using an absorbed APEC model with fixed Galactic
absorption, abundance fixed at $0.3\Zs$ and local backgrounds 
from Table \ref{tab:lbgreg},  we
find the temperature of the gas in this region to 
be $19_{-4}^{+6}$~keV ($\chi^2/{\rm dof} = 165/174$). This region 
corresponds to the high temperature ridge between the subcluster and 
primary cluster in the right panel of Figure \ref{fig:resid}. A surface 
brightness profile centered on the primary cluster cD galaxy taken in 
the sector subtending the angles from $198^{\circ}$ to $258^{\circ}$ with 
semi-major (-minor) axes and position angle of~$56''$ ($15''$) and 
$138^{\circ}$, respectively, traces the shape of the decrement as well 
as the proposed western shock front. The profile indicates increased 
surface brightness over the edge. See Figure \ref{fig:west_sfell} for 
the profile with shock region in red, and right panel of 
Figure \ref{fig:resid} for the profile's location and shock region 
enclosed in black. High temperatures, high surface brightness, 
and a significant SZ decrement in this region suggest that the edge 
is the western shock front of the subcluster's Mach cone. 

The opening angle $\mu$ of the Mach cone can be
calculated by \citep{markevitch2007}
\begin{eqnarray}
\label{eq:mach}
sin\left(\mu\right)=\mbox{$\mathcal M^{-1}$}.
\end{eqnarray}
For our calculated eastern Mach number
of $1.25 \pm 0.08$, we predict an opening angle of 
$\mu=53.1 \pm 0.1$\,degrees. Assuming the Mach cone is defined as the 
angle between the eastern shock edge (see Table \ref{tab:sfparam}) and
the SZ decrement, we measure an opening angle $\mu\approx50$ degrees, 
in good agreement with the predicted value.

\begin{figure}[b!]
\begin{center}
\includegraphics[width=0.475\textwidth]{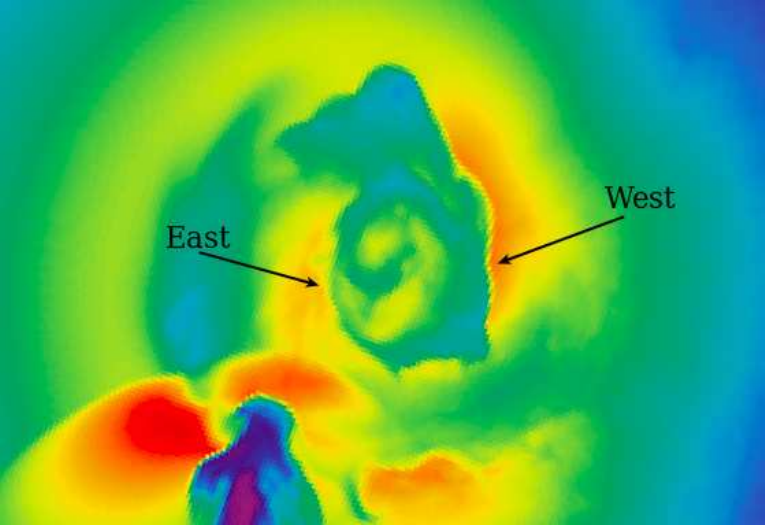}
\caption{A simulated spectroscopic-like projected temperature map 
from \citet{zuhonesim} and \citet{johnson2012}. We label the same 
edges as \citet{johnson2012}.
}
\label{fig:johnson_orig}
\end{center}
\end{figure}

\begin{figure*}[t!]
\begin{center}
\includegraphics[width=0.475\textwidth]{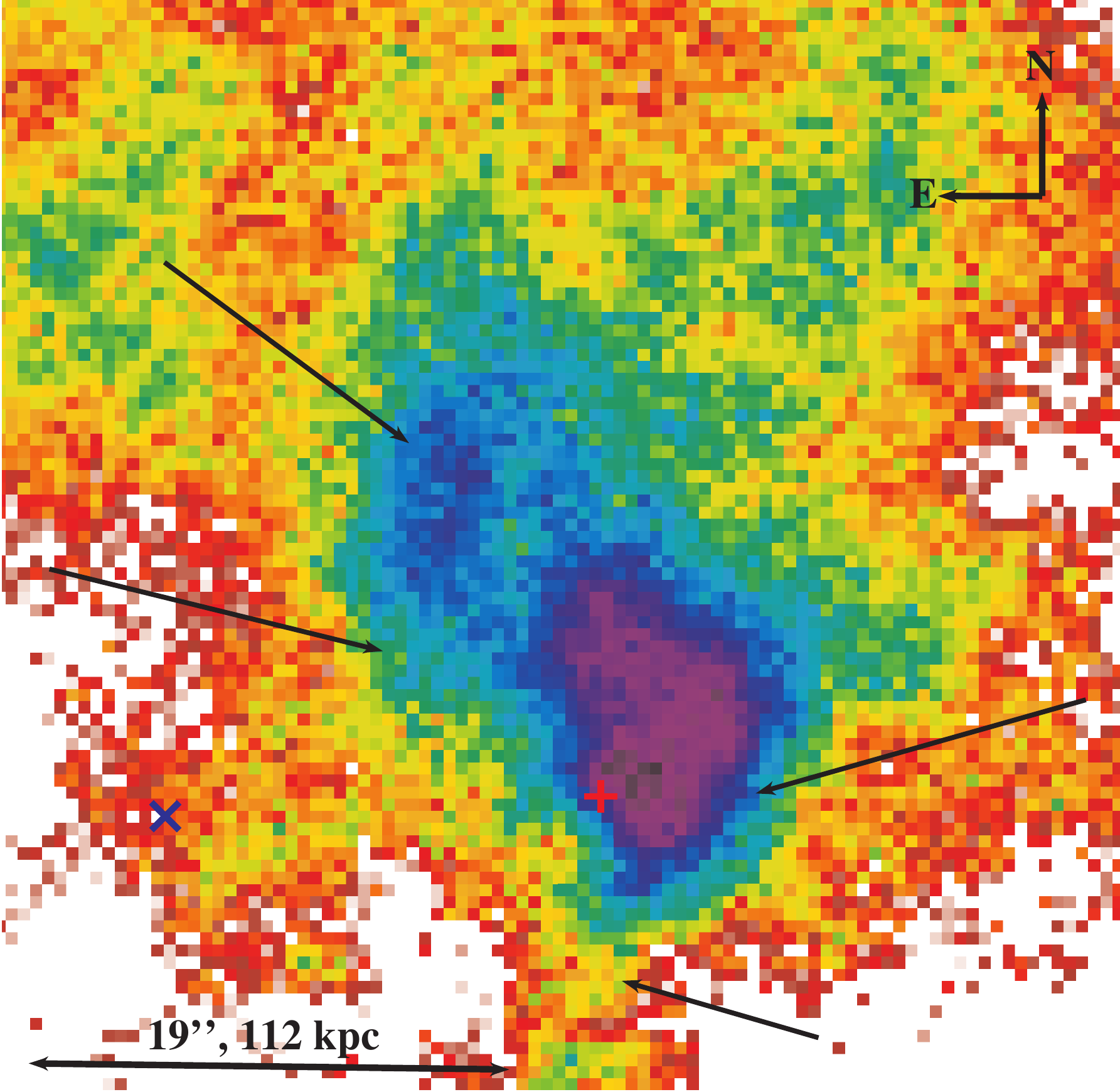}
\includegraphics[width=0.475\textwidth]{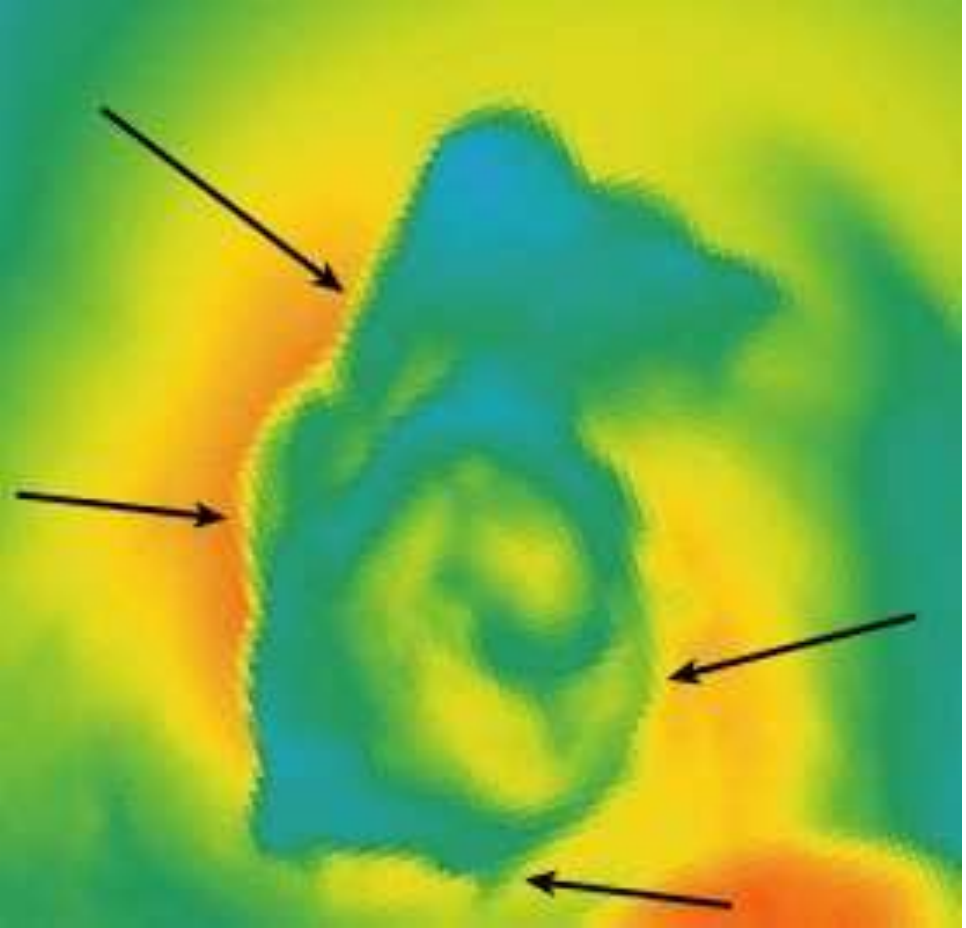}
\caption{In the left panel we show the projected temperature map with
  cold fronts highlighted. To the right we display the same simulated
  projected temperature map 
  as in Figure \protect\ref{fig:johnson_orig} flipped about its N-S
  axis.  We label edges that appear to match those 
  found in the data.
}
\label{fig:temp_sim}
\end{center}
\end{figure*}

\section{Cluster Merger History: A Comparison to Simulations}
\label{sec:sim}

To constrain the cluster's merger history, it is important to compare
our observations with simulations. In Figure \ref{fig:johnson_orig} we 
present the same simulation and labels shown in \citet{johnson2012}, 
who argue that the subcluster is on its second pass around the primary 
cluster. The simulation initial conditions are described in
\citet{zuhonesim} and employ a mass ratio of the primary cluster to 
the subcluster of $10$ and an impact parameter of~$\sim 1$~Mpc. The 
subcluster is on a clockwise orbit about the primary cluster, such that 
it first passes the primary cluster in the west. The simulation
captures the subcluster's position and the shock fronts well, in 
comparison with the X-ray data. However, the cold front locations 
are not consistent with our observations. In the simulations, the 
eastern edge is closer 
to the primary BCG, suggesting that it formed later 
than the western cold front, but our observations definitively 
show that the western cold front is closer to the primary BCG. Thus
our data suggest that the western cold front formed more recently than 
the eastern cold front. This leads us to 
consider the possibility that more than 1 subcluster has perturbed RXJ1347. 

In the right panel of Figure \ref{fig:temp_sim}, we display the same 
simulation but flipped about the N-S axis such that the subcluster
perturber in the simulation is now 
in a counterclockwise orbit, first passing to the east of the primary 
cluster. The merger is in the plane of the sky, and we compare our 
temperature map to this simulation in the left panel. If we ignore 
the location of the subcluster in the simulation (now in the
southwest), we find many similarities between our deep {\it Chandra} 
observations and the simulation. The four cold fronts forming a 
clockwise spiral in the observations can be located in the
simulation. We denote these fronts with arrows in both the 
simulation and the temperature map. The fronts also form a 
clockwise spiral in the simulation. The fan found in the X-ray 
observations, and most prominently seen in our residuals image, 
is similar to the cool gas forming part of the spiral that extends 
to the north and northwest in the simulation. 

As the subcluster passes the primary cluster, it transfers angular
momentum \citep{ascasibar2006}, so we expect the observed clockwise 
spiral, seen  in Figure \ref{fig:xray} and confirmed by our cold front
analysis, to have been caused by interaction with a subcluster on a 
counterclockwise orbit around the primary cluster. The clockwise
spiral in the flipped simulation supports this 
scenario. This discrepancy between the observed subcluster to 
the southeast moving in a clockwise orbit, and the clockwise sloshing spiral in the primary cluster produced in simulations by multiple passes of a perturber moving in a counterclockwise orbit, suggests that the observed subcluster may be on its first pass through the primary cluster, while the clockwise sloshing spiral was caused by earlier encounters with a different subcluster moving in a counterclockwise orbit. In this scenario, the observed subcluster is being ram-pressure stripped while shock heating the cluster gas. The earlier perturber, whose interaction caused the sloshing spiral, may no longer be visible in X-rays because nearly all of its gas has already been stripped. RXJ1347 is in a $\sim20$~Mpc filament, 
so it would not be surprising if it had previously interacted with 
another object in this structure \citep{filament}. 

We searched the literature for evidence in other wavelengths of a second merging subcluster. \citet{lu2010} find a massive cluster located $\sim 7$ Mpc southwest of RXJ1347 (RXJ1347-SW) with $\sigma=780 \pm 100$ km s$^{-1}$, $M_{200}=3.4^{+1.4}_{-1.1} \times 10^{14}\,M_{\odot}$, and mean redshift $0.4708 \pm 0.0006$. RXJ1347-SW has a large relative radial velocity of $4000$ km s$^{-1}$. \citet{lu2010} argue the likelihood RXJ1347-SW is falling into RXJ1347 is low. Nonetheless, the concentration of galaxies between RXJ1347 and RXJ1347-SW in velocity space implies the two clusters are physically linked by the same large scale filament.

The recent strong lensing analysis of \citet{schmidt}, using redshift measurements from the Cluster Lensing and Supernova survey \citep[CLASH,][]{postman}, requires two additional perturbers compared to previous lensing analyses \citep{halkola,bradac2008} to model the observed arc geometry. One is a dark halo aligned with a faint $z_{phot}>0.7$ object and not associated with the primary cluster, and the second is a massive ($\sigma_v\sim763$ km s$^{-1}$) concentration of cluster galaxies to the NW. While the modeled NW mass component is intriguing as a possible subcluster remnant, it does not appear coincident with any bright cluster galaxy, and no redshift is given. Thus, more work is needed to verify the NW component's existence and association with the primary cluster, as well as to resolve discrepancies between the CLASH redshift measurements of the lensed systems and those of previous work \citep{halkola,bradac2008}.

Another possible 
scenario is that the orbit of the observed subcluster is not 
in the plane of the sky and that the observed clockwise orientation of 
the edges may be due to projection effects. While we consider this
latter scenario unlikely, it cannot be ruled out without better 
simulation studies. 

\section{CONCLUSIONS}
\label{sec:conclude}

With~$2.5$~times the exposure of previous analyses, we are able to
study the gas hydrodynamics of RXJ1347 and its merging subcluster in 
unprecedented detail. We briefly summarize our key results.
\begin{itemize}
\item{A series of four cold fronts west, southeast, east, and
    northeast located at~$5\farcs 85^{+0.04}_{-0.03}$,
 $7\farcs 10^{+0.07}_{-0.03}$, $11\farcs 5^{+1.3}_{-1.2}$, 
and $16\farcs 7^{+0.3}_{-0.5}$~from the primary cluster's cD galaxy, respectively, 
forms a clockwise spiral, suggesting the merger is in the plane of the
sky. The west and east cold fronts correspond to those found 
by \citet{johnson2012}.}  
\item{We measure a subcluster $0.5-2$\,keV luminosity of 
$L_x= 1.3 \times 10^{44}$\ergs and infer a lower bound on the total subcluster 
mass of $3.3 \times10^{14}$~M$_\odot$.}
\item{We identify the shock from the supersonic infall of the
    subcluster and measure a density ratio across the shock 
 of $1.38^{+0.16}_{-0.15}$~, corresponding to a Mach number of
 $1.25\pm0.08$ and shock velocity of~$2810^{+210}_{-240}$~km~s$^{-1}$. 
 The measured opening angle of the Mach cone of~$\sim50$~deg is consistent with this 
shock interpretation.}
\item{The subcluster's baryonic gas has been stripped from the dark 
matter peak and leaves a tail behind the subcluster. Asymmetric
entropy contours and cooler temperatures extending through the
subcluster and tail indicate the subcluster core is being shredded.}
\item{We suggest that the most likely explanation for the observed X-ray features is that the southeast subcluster is on its first passage on a clockwise orbit through the cluster, shock heating the cluster gas as it infalls, while the clockwise sloshing spiral observed in the primary cluster was caused by earlier encounters with a second subcluster moving in a counterclockwise orbit. Better simulations and further observations in other wavelengths are needed to search for remnants of this second perturber and test this scenario.}
\end{itemize}

\acknowledgements

This work was supported in part by NASA CXC grant GO2-13148X
and the Smithsonian Institution. CK gratefully acknowledges support
from the Smithsonian Institution Minority Awards Program. This
research has made use of data obtained from the Chandra Data Archive 
and the Chandra Source Catalog, and software provided by the Chandra 
X-ray Center (CXC) in the application 
packages CIAO and Sherpa. This work has made use of the NASA/IPAC 
Extragalactic Database (NED) which is operated by the Jet Propulsion 
Laboratory, California Institute of Technology,  under contract with 
the National Aeronautics and Space Administration. We would like to 
thank Maxim Markevitch for helpful discussions and use of his
multivariate~$\chi^2$~minimization code for fitting the surface
brightness profiles, Brian Mason for use
of his SZ contours, Jeremy Sanders for help with his GGM filtering technique, Sherry Suyu for useful strong lensing discussions, and the anonymous referee for exceptionally helpful comments.  

\noindent{\it Facilities:} CXO (ACIS-I) 

\vfill
\eject
\bibliography{paper}
\end{document}